\documentclass[twoside]{JHEP3}
\usepackage[english]{babel}
\usepackage{amssymb}
\usepackage{amsfonts}
\usepackage{amsmath}
\usepackage{graphicx}
\usepackage{epsfig}
\usepackage{cite}
\def\erf#1{(\ref{#1})} 
\newcommand{\cA}{{\cal A}}  \newcommand{\cB}{{\cal B}}

\newcommand{\bk}{{\mathbf k}}  \newcommand{\bq}{{\mathbf q}}
\newcommand{\bx}{{\mathbf x}}

\def\bsxi{\boldsymbol{\xi}}

\def\wn{{\mathfrak{w}}}  \def\qn{{\mathfrak{q}}}

\newcommand{\be}{\begin{equation}} \newcommand{\ee}{\end{equation}}
\newcommand{\bea}{\begin{eqnarray}} \newcommand{\eea}{\end{eqnarray}}
\newcommand{\beann}{\begin{eqnarray*}}  \newcommand{\eeann}{\end{eqnarray*}}
\newcommand{\bfig}{\begin{figure}} \newcommand{\efig}{\end{figure}}
\newcommand{\ba}{\begin{array}} \newcommand{\ea}{\end{array}}
\newcommand{\bcen}{\begin{center}} \newcommand{\ecen}{\end{center}}
\newcommand{\btab}{\begin{tabular}} \newcommand{\etab}{\end{tabular}}

     \def\sign{\operatorname{sign}}
\renewcommand{\Re}{\mathop{\rm Re}}   \renewcommand{\Im}{\mathop{\rm Im}}
\newcommand{\vev}[1]{\left\langle{#1}\right\rangle}

\newcommand{\dd}{{\rm d}}
\newcommand{\e}{{\rm e}}

  \def\Nfour{{\cal N}\!=\!4}

\newtheorem{Proposition}{Proposition}[section]

\newtheorem{Theorem}{Theorem}[section]
\newtheorem{Lemma}{Lemma}[section]
\newtheorem{Corrolary}{Corrolary}[section]

\newcommand{\bp}{\begin{Proposition}}	\newcommand{\ep}{\end{Proposition}}
\newcommand{\bt}{\begin{Theorem}}	\newcommand{\et}{\end{Theorem}}
\newcommand{\bl}{\begin{Lemma}}		\newcommand{\el}{\end{Lemma}}
\newcommand{\bc}{\begin{Corrolary}}	\newcommand{\ec}{\end{Corrolary}}
\title{Hydrodynamics and beyond in the strongly coupled {\boldmath $\Nfour$} plasma}
\author{Irene Amado,${}^a$ Carlos Hoyos,${}^b$ Karl Landsteiner${}^a$ and Sergio Montero${}^a$\\
  ${}^a$Instituto de F\'{\i}sica Te\'orica IFT-UAM/CSIC\\ 
  ~\,Universidad Aut\'onoma de Madrid\\
  ~\,E-28049 Madrid, Spain\\
  ~\,E-mail: \email{Irene.Amado, Karl.Landsteiner, Sergio.Montero@uam.es}\\
  ${}^b$Department of Physics,
  Swansea University\\
  ~\,Swansea, SA2 8PP, UK\\
  ~\,E-mail: \email{C.H.Badajoz@swansea.ac.uk}}
\abstract{We continue our investigations on the relation between hydrodynamic and higher quasinormal modes in the AdS black hole background started in arXiv:0710.4458 [hep-th]. As is well known, the quasinormal modes can be interpreted as the poles of the retarded Green functions of the dual $\Nfour$ gauge theory at finite temperature. The response to a generic perturbation is determined by the residues of the poles. We compute these residues numerically for energy-momentum and R-charge correlators. We find that the diffusion modes behave in a similar way: at small wavelengths the residues go over into a form of a damped oscillation and therefore these modes decouple at short distances. The sound mode behaves differently: its residue does not decay and at short wavelengths this mode behaves as the higher quasinormal modes. Applications of our findings include the definition of hydrodynamic length and time scales. We also show that the quasinormal modes, including the hydrodynamic diffusion modes, obey causality.}
\preprint{IFT-UAM/CSIC-08-28} 

\begin{document}
%
\section{\label{sec:intro}Introduction}
In recent years a new paradigm concerning the high temperature behaviour of QCD has been established: the strongly coupled Quark-Gluon Plasma (sQGP). Experimental results from heavy ion collisions at RHIC indicate that QCD at temperatures around $2T_c$ is strongly interacting, in spite of being in a deconfined phase, and thus rendering perturbative computations not suitable for describing it. While static properties of strongly coupled gauge theories at finite temperature can be readily analyzed on the lattice, the study of out-of-equilibrium phenomena faces considerable difficulties. In the last years the AdS/CFT correspondence \cite{Maldacena:1997re, Gubser:1998bc, Witten:1998qj} has emerged as a useful tool to understand analytically the dynamics of non-Abelian gauge theories in the strongly coupled plasma phase.

According to AdS/CFT, the asymptotically AdS black hole is dual to the plasma phase of the  strongly coupled $\Nfour$ gauge theory \cite{Gubser:1996de, Klebanov:1996un, Witten:1998zw}. The line element is 
\begin{equation}\label{eq:AdSBH}
\dd s^2 = \frac{r^2}{L^2} \left(- f(r) \dd t^2 + \dd\bx^2 \right)  + \frac{L^2}{r^2}\frac{\dd r^2}{f(r)} ~,
\end{equation}
with $f(r)= (1-r_0^4/r^4)$. The Hawking temperature $T_H = r_0/\pi L^2$ is the temperature $T$ of the dual field theory. Real-time correlation functions in the thermal gauge theory can be computed using classical solutions of fields living in the gravity dual \cite{Son:2002sd}. If a field is excited in the presence of a black hole, the energy of the fluctuation will be lost inside the horizon and eventually the final state will be a larger black hole with no fluctuations. This process is described by the quasinormal spectrum, that was first computed for black holes in asymptotically Anti de Sitter (AdS) spacetimes in \cite{Horowitz:1999jd}. Quasinormal modes (at fixed momentum) exist only for a discrete set of complex frequencies that can be identified with the poles of the retarded Green functions or resonances in the dual field theory \cite{Birmingham:2001pj, Starinets:2002br, Nunez:2003eq}. Therefore, quasinormal modes describe dissipation processes in the plasma. 

The retarded Green functions depend on frequency $\omega$  and momentum $\bq$. The quasinormal modes can be understood as solutions of  $(G_\mathrm{R})^{-1} =0$ in the complexified $\omega$ plane. However, one can also ask for solutions in the complexified momentum plane. Again one finds a discrete spectrum of complex momentum modes (keeping the frequency real) whose imaginary part can be interpreted as the inverse absorption length \cite{Amado:2007pv}. We will also consider these complex momentum modes and their residues in this paper.

Hydrodynamic modes like diffusion, shear or sound modes are also described by quasinormal modes in the gravity dual. They are special cases of quasinormal modes whose frequencies vanish in the zero momentum limit \cite{Policastro:2002se, Policastro:2002tn}. One of the most interesting results of this AdS black hole hydrodynamics has been the derivation of a universal bound for the shear viscosity to entropy ratio $\frac{\eta}{s} \geq \frac{\hbar}{4\pi k_B}$ \cite{Policastro:2001yc, Kovtun:2004de}. It has also been argued that this value is relevant to the description of heavy ion phenomenology at RHIC \cite{Muller:2006ee}. The shear viscosity is a hydrodynamic transport coefficient governing the momentum diffusion through the medium. One way to derive it is to compute the lowest quasinormal frequency in the retarded two-point correlator in the vector channel of the stress tensor. 

The hydrodynamic approximation can be interpreted as a low energy effective theory where all the degrees of freedom except the hydrodynamic modes have been integrated out. It is then possible to make a systematic expansion in higher derivative terms, whose coefficients depend on the microscopic physics. The first order terms correspond to the usual hydrodynamic equations, while second order terms take into account the finite delay in the response of the system. Second order hydrodynamics is important since the relativistic first order formalism leads to the well-known problem of acausal signal propagation. As we will review in appendix \ref{sec:frontveloc}, the speed of signal propagation is determined by the so called front velocity $v_{\rm}=\lim_{\omega\rightarrow \infty} \frac{\omega}{q^R}$, with $q^R$ the real part of the complex momentum mode. Diffusion equations as they arise typically in first order hydrodynamics do not have a finite front velocity and therefore allow acausal behaviour. In second order hydrodynamics the non-zero delay of the response is encoded in a relaxation time constant and can restore causality \cite{Israel:1976tn, Mueller, Stewart, Israel:1979wp}. The latter have been recently computed for conformal theories using the AdS/CFT correspondence \cite{Heller:2007qt, Baier:2007ix, Bhattacharyya:2008jc, Natsuume:2007ty }. However, even second order hydrodynamics cannot simply be extended to arbitrary short wavelengths or high frequencies and thus a front velocity smaller than the speed of light in second order hydrodynamics is not guaranteed, since inevitably the hydrodynamic description breaks down at small wavelengths. Thus the question arises how causal signal propagation is guaranteed in the $\Nfour$ plasma at strong coupling. We will show numerically that the hydrodynamic mode approaches the front velocity $v_{\rm F}=1$ and argue that this is indeed the exact value for all quasinormal modes including the hydrodynamic diffusion modes.

As an effective theory, the hydrodynamic approximation will be valid up to energies of the order of the microscopic scale, in this case given by the temperature $T$. At weak coupling hydrodynamic modes are collective excitations made up by the collective motion of the hard particles. Therefore they actually should decouple at short wavelengths. In standard perturbative finite temperature field theory and in the HTL approximation collective excitations do indeed appear already in the two-point correlators of the fundamental fermion and gauge fields. At weak coupling one distinguishes a soft scale of order $gT$ from a hard scale $T$. Above $T$ the hard partons are the quarks and gluons of the elementary fields, at the soft scale there are quasiparticles, dressed quarks and gluons and collective excitations, i.e. the plasmon and plasmino modes \cite{LeBellacBook, Kraemmer:2003gd, Blaizot:2001nr}.  They show up as poles in the retarded two-point functions but can be distinguished by the behaviour of the residues at momenta at or above the hard scale $T$: the residues of the collective excitations decay exponentially signalling their decoupling at short wavelengths. What can we expect then for the quasinormal modes? In fact the high frequency and short wavelength limit can be interchanged with the $T\rightarrow 0$ limit \cite{Starinets:2002br}. Going to zero temperature an infinite number of quasinormal modes come together and open up a branch cut, characteristic for the $T=0$ retarded correlators. Therefore, the residues of almost all quasinormal modes should not decay in this limit. However, a finite number might still decouple, and indeed this is what we find for the residues of quasinormal modes representing diffusive behaviour at long wavelengths.

In this work we extend previous analysis of the residues of the quasinormal modes of the R-current \cite{Amado:2007yr} to energy-momentum tensor correlators. In section \ref{sec:linresponse} we recall linear response theory and how it is related to hydrodynamic behaviour. We study the regime of validity of the hydrodynamic approximation and define a lower bound in length and time scales. In section \ref{sec:resonances} we discuss how the retarded Green function can be expressed as infinite sums over the poles at the quasinormal frequencies or complex momenta. We point out that these sums do not converge and analytic terms are needed to regularize the sums. These analytic terms should not be confused with contact terms arising from the action.

In section \ref{sec:vector} we recall and extend the results of ref. \cite{Amado:2007yr} on R-charge correlators. In particular we compute the residues of the complex momentum modes whose imaginary parts give the inverse absorption lengths \cite{Amado:2007pv} and study how well can the real and imaginary parts of the retarded Green functions be approximated by the contributions of the lowest quasinormal modes.

In section \ref{sec:metric} we study quasinormal frequencies, complex momentum modes and their residues for stress tensor correlators. We find that the residues of the shear mode quasinormal frequency go over into a form similar to a damped oscillation at short wavelengths. Numerically we find that it decouples for short wavelengths. The behaviour of the quasinormal frequencies in the shear channel are very similar to the diffusion channel for the R-charge correlators. The hydrodynamic mode crosses the imaginary parts of the other (low) quasinormal frequencies roughly at the locations of the zeroes in the shear mode residue, signalling a breakdown of the effective field theory based on the shear mode alone. We also find that the residues in the sound channel show a different behaviour: the lowest mode, the sound mode, does not decouple at short wavelengths. It becomes similar to a higher quasinormal mode. Again we study how well the spectral functions can be approximated by summing only over the lowest quasinormal modes. In addition we also compute the residues in the shear and sound channels of the complex momentum eigenvalues for fixed real frequencies. This is important in order to compute (numerically) the front velocity. We find that the front velocity in all modes, even the hydrodynamic one, approaches unity at high frequency and give an argument that this is indeed enforced by the Lorentz symmetry of the underlying $T=0$ theory.

In appendix \ref{sec:method} we collect technical details on how we compute the retarded Green functions, their poles and residues. In appendix \ref{sec:zeroes} we show how the location of the zeroes in the residues of R-charge diffusion and shear mode are determined by algebraic equations arising from the recursion relations of the corresponding Heun equations. Finally in appendix \ref{sec:frontveloc} we recall the definition of the front velocity.

\section{\label{sec:linresponse}Hydrodynamic scales and linear response theory}
We consider a medium in thermal equilibrium. To study the effect of small external perturbations it is enough to use linear response theory when the energy of the perturbation is negligible compared to the total energy of the system. In this linear approximation, the response of a field $\Phi$ to a perturbation represented by an external source $j(t,\bx)$ is 
\begin{equation}\label{eq:linear_response}
\vev{\Phi(t,\bx)} =-\int \dd\tau\,\dd^3\bsxi \,G_{\rm R}(t-\tau, \bx-\bsxi) \,j(\tau,\bsxi) ~,
\end{equation}
where $G_{\rm R}$ is the retarded two-point correlation function,
\begin{equation}
G_{\rm R} (t-\tau,\bx-\bsxi) = -i \,\Theta(t-\tau) \vev{\left[ \Phi(t,\bx) , \Phi(\tau, \bsxi)\right]}\,.
\end{equation}
Using the Fourier transforms of the Green function and the source this is rewritten as
\begin{equation}
\vev{\Phi(t,\bx)} = - \int \frac{\dd\omega}{2\pi} \frac{\dd^3\bq}{(2\pi)^3} \,\e^{-i\omega t+i \bq \bx} \,\widetilde{G}_{\rm R}(\omega, \bq )\, \tilde{\jmath}(\omega,\bq) ~.
\end{equation}
We can now make the analytical continuation to the complex $\omega$ plane and use Cauchy's theorem to evaluate this integral by closing the contour on the lower half of the complex $\omega$ plane for $t>0$, thus picking up the contributions from the poles of the retarded propagator, i.e. the frequencies of the quasinormal modes in the AdS/CFT correspondence. At this point we assume that the retarded Green function is analytic in the upper half of the complex frequency plane and that its only singularities are single poles in the lower half plane. This is indeed the case for the holographic retarded two-point functions. We also assume that the source does not introduce new non-analyticities in the lower half plane. The retarded propagator is therefore of the form
\begin{equation}
\widetilde{G}_{\rm R}(\omega, \bq ) \simeq  \sum_{\rm poles} \frac{ R_n(\omega,\bq) }{\omega - \omega_n (\bq)} ~,
\end{equation}
where $\omega_n := \Omega_n - i \Gamma_n$ and $R_n$ are the residues of the retarded propagator evaluated at the poles. We should emphasize that this is a formal expression since the infinite sum does not necessarily converge; we will come back to this issue in section \ref{sec:resonances}. Doing a partial Fourier transform in space, the response of the system to a perturbation is now expressed by a sum over the poles 
\begin{equation}\label{eq:linear_response_QNM}
\vev{\Phi(t,\bq)} = i \theta(t) \sum_{n} R_n(\omega_n(q),\bq) \,\tilde\jmath(\omega_n(\bq), \bq) \,\e^{ -i \Omega_n(\bq) t - \Gamma_n(\bq) t} ~.
\end{equation}
The location of the quasinormal modes determine their frequency and damping and their residues determine how much each mode contributes to the response. The quasinormal mode spectrum of the planar AdS black hole is well known by now \cite{Starinets:2002br,Friess:2006kw, Musiri:2005ev, Siopsis:2004up}. For each channel corresponding to a gauge invariant operator in the dual gauge theory there is an infinite tower of quasinormal modes. Since for holographic duals of gauge theories there exist infinitely many quasinormal modes the precise form of the response also depends on the form of the source. There is however an exception to this: the extreme long-wavelength modes of conserved charges such as energy and momentum. Perturbations that induce a change in the overall charge have to excite the special hydrodynamic modes. These are modes whose quasinormal frequencies obey $\lim_{\bq\rightarrow 0} \omega_{\rm H}(\bq) =0$. This behaviour guarantees that the integrated response $\int \dd^3\bx \vev{\Phi(t,\bx)}$ is time independent, as it has to be for a conserved charge. Notice however that there is no fundamental reason that would forbid neutral perturbations to dissipate away primarily in the higher quasinormal modes, i.e. the infinite tower of modes with non vanishing zero momentum limit. It is also interesting to think about what the presence of the infinitely many quasinormal modes mean for the formulation of an initial value problem. Let us assume for the moment that there would be only one quasinormal mode, e.g. a diffusion mode in a conserved charge. The time development for $t>0$ would then be completely determined by specifying the expectation value of the field at $t=0$, i.e. $\vev{\Phi(0,\bx)} = \Phi_0(\bx)$. Indeed, cutting the sum in (\ref{eq:linear_response_QNM}) after the first term, the hydrodynamic term in this case, leaves us with a one-to-one relationship between the initial value  and the source $\widetilde\Phi_0(\bq)= R(\bq) \,\tilde\jmath(\bq)$. However, in our case there are infinitely many quasinormal modes and therefore one has to specify not only the field at $t=0$ but also an infinite number of its time derivatives in order to be able to compute the source from the initial condition. Strictly speaking the system becomes non-Markovian due to the presence of the infinitely many quasinormal modes: it remembers (at least for short times) the history of how it reached a certain state. Putting it another way, we can define a system to be in local thermal equilibrium if it has no memory, i.e. the response is completely dominated by the hydrodynamic mode.

We will concentrate on simple forms of source perturbations. We imagine that the source acts only over a time interval $\Delta t$. Such a perturbation will excite a significant number of quasinormal modes for small $\Delta t$ and in the limit where $\Delta t \rightarrow 0$ it will excite all quasinormal modes with equal weight. We can then assume a source of the form $ j(t,\bx) = \delta(t) \cos(\bq \bx)$. 

At small values of the momentum, the hydrodynamic mode dominates the long time behaviour since it has the smallest imaginary part. In fact, we can define the hydrodynamic time scale, from which on the hydrodynamic mode dominates and the hydrodynamic approximation is good, by demanding that the response in the hydrodynamic mode equals the response in the first quasinormal mode at the time $\tau_{\rm H}$. We can estimate this time scale using \erf{eq:linear_response_QNM} 
\begin{equation}\label{eq:hydro_time}
\tau_{\rm H} = \frac{\log|R_{\rm H} \,\tilde\jmath_{\rm H}| -\log|R_1 \,\tilde\jmath_1|}{ \Gamma_{\rm H} - \Gamma_1} ~,
\end{equation}
where we have written $\tilde\jmath_n = \tilde\jmath(\omega_n(\bq),\bq)$. The hydrodynamic description will be trustable for times larger than $\tau_{\rm H}$, whereas for shorter times the contribution of higher modes must be taken into account. 

Now  we switch the roles of time and one spatial coordinate and choose a periodic perturbation localized in space, over an interval $\Delta x$ in the $x$ direction. In the limit $\Delta x \to 0$ we assume a source of the form $j(t,\bx) = \delta(x) {\rm exp}[-i(\omega t - \bk_{\perp} \bx_{\perp})]$. From \erf{eq:linear_response}, the response of the system takes the form
\begin{equation}
\vev{\Phi(t,\bx)} = -\frac{1}{2\pi} \, \e^{-i(\omega t - \bk_{\perp} \bx_{\perp})} \int \dd q \, \e^{iqx} \,\widetilde{G}_{\rm R}(\omega, \bk_\perp, q ) ~.
\end{equation}
Such a perturbation has the form of a plane wave in the perpendicular directions $\bx_{\perp}$. We have assumed that the perturbation started far in the past such that the system has reached a stationary state. We will also assume that it is no further modulated in the $\bx_{\perp}$-directions, i.e. we set $\bk_{\perp}=0$. By Cauchy's theorem we can evaluate this integral by closing the contour on the upper half of the complex $q$ plane for $x>0$ (on the lower-half complex $q$ plane for $x<0$). The contributions from the poles of the retarded propagator in the complex $q$ plane correspond to complex wavenumber modes in the AdS/CFT correspondence, whose imaginary part determine the absorption lengths of perturbations in the plasma \cite{Amado:2007pv, Hoyos:2006gb}. We consider again that the only singularities of the retarded propagator are single poles in the complex momentum plane, so
\begin{equation}
\widetilde{G}_{\rm R}(\omega, q ) \simeq \sum_{\rm poles} \frac{ R_n'(\omega,q ) }{q - q_n (\omega)} ~,
\end{equation}
where $q_n=q_n^R+iq_n^I$. The response of the system to a perturbation localized in space as a sum over the poles reads
\begin{equation}\label{eq:linear_response_CMM}
\langle \Phi(t,\bx) \rangle = -i \,\sign(x) \,\e^{-i\omega t} \sum_{n} R_n'(\omega,q_n(\omega)) \,\e^{ i q_n^R(\omega) x - q_n^I(\omega) x} ~.
\end{equation}
By parity symmetry $x\rightarrow -x$, if there is a pole at $q_n=q_n^R+iq_n^I$, then also $q=-q_n$ has to be a pole, so the poles of the retarded Green function lie in the first and third quadrants of the complex $q$ plane for $x>0$ and $x<0$, respectively. On the supergravity side this has been indeed proved in \cite{Amado:2007yr}.

Analogously to what we did before, we can define a hydrodynamic length scale. From \erf{eq:linear_response_CMM} we find
\begin{equation}\label{eq:hydro_length2}
\ell_{\rm H} = \frac{{\rm log}|R_{\rm H}'|- {\rm log}|R_1'|}{\Im q_{\rm H} - \Im q_1} ~.
\end{equation}
For distances from the origin of the perturbation larger than $\ell_{\rm H}$, the hydrodynamic mode will dominate the response of the system. We emphasize that the definition of the hydrodynamic length scale (\ref{eq:hydro_length2})  applies only if the sources are localized in space. For a more general source, the Fourier transform evaluated at the complex momentum mode $\tilde\jmath(\omega,\bq_n(\omega))$ will enter in the definition of $\ell_{\rm H}$ in an analogous way to (\ref{eq:hydro_time}).

As we have seen the response and the hydrodynamic scales depend crucially on the knowledge of the residues of the retarded correlators. In the following we will determine them numerically and study such hydrodynamic scales.

\section{\label{sec:resonances}Including higher thermal resonances}
The hydrodynamic mode provides a good description of the system at very long wavelengths $\omega, q\ll T$. As we have argued, for shorter times or distances the holographic computation shows that higher modes start to be relevant. It is an interesting question to see if a description in terms of the hydrodynamic mode plus a few thermal resonances gives a reasonable good model of the plasma up to frequencies of the order of the temperature $\omega \sim T$. What we mean by thermal resonances are the quasinormal modes found in the holographic dual, that describe the dissipation of gauge-invariant configurations. 

Due to conformality, the properties of the ${\cal N}=4$ plasma rescale trivially with the temperature $T$. A possible way to introduce a non-trivial temperature dependence would be to compactify the theory in a three sphere of radius $R$. The compactification breaks conformal invariance, so there is a non-trivial dependence on $R\,T$. Then, the physics of the plasma in flat space $R\to \infty$ can also be recovered in the infinite temperature limit $T\to \infty$ \cite{Witten:1998zw}. Computations of the quasinormal mode spectrum at lower temperatures $R\,T \geq 1$ \cite{Horowitz:1999jd, Konoplya:2002zu} show that the lowest modes are long lived enough to have a good quasiparticle interpretation. Other computations in the context of flavour branes, where high and low temperature is given in terms  of the quark mass $m_q/T$, also show similar results for the quasinormal spectrum of mesons \cite{Hoyos:2006gb, Hoyos:2007zz, Myers:2007we, Paredes:2008nf}. This suggests that for more realistic plasmas, e.g. non-conformal and maybe closer to the intermediate temperature regime between the deconfinement transition and the free gas limit than the simpler models just mentioned, a description in terms of bound states could be appropriate, although usually coloured bound states are considered \cite{Shuryak:2003ty, Brown:2003km, Shuryak:2004tx}. In spirit, this approach is similar to the description of QCD correlators at low energies using the lowest states of the meson spectrum.

From the linearized AdS/CFT computation, we know that the retarded correlators of channels with hydrodynamic modes have the general form
\begin{equation}\label{eq:greenqnms}
\widetilde{G}_{\rm R}(\omega,q)=\sum_{i=1}^{n_{\rm H}} \frac{R_{\rm H}^{(i)}}{\omega-\Omega_{\rm H}^{(i)}} + \sum_{n=1}^\infty \frac{R_n}{ \omega-\omega_n} + \frac{-R_n^*}{\omega+\omega_n^*}+\cA(\omega,q) ~,
\end{equation}
where $\Omega_{\rm H}$, $\omega_n$ are the hydrodynamic and quasinormal poles, $R_{\rm H}$, $R_n$ their residues and $n_{\rm H}$ is the number of hydrodynamic modes. For the shear and diffusive channels $n_{\rm H}=1$, while for the sound channel $n_{\rm H}=2$ and $\Omega_{\rm H}^{(2)}=-\Omega_{\rm H}^{(1)\,*}$, $R_{\rm H}^{(2)}=-R_{\rm H}^{(1)\,*}.$\footnote{There exist of course also channels without any hydrodynamic behaviour $n_H=0$, such as scalar field perturbations. We will not consider these channels in this paper.} The term $\cA$ denotes possible terms analytic in frequency and momentum. In principle they look like the contact terms that can arise from Ward identities or some choice of renormalization scheme, in which case they could be absorbed by a finite number of counterterms. However, their origin is different and we will see that they are necessary in order to give a consistent Green function, the main reason being the infinite sum over quasinormal modes. 

Let us concentrate on a particular example, the density-density correlator of the R-current. In \cite{Myers:2007we} the explicit form of the current-current correlator at zero momentum was found ($\wn:=\omega/(2 \pi T)$),
\begin{equation}
G_{ij}(\wn) = \delta_{ij}\frac{N^2 T^2}{8} \Bigg\{ i \wn  + \wn^2 \left[\psi\left( \frac{(1-i)}{2}\wn\right) + \psi\left( \frac{-(1+i)}{2}\wn\right) \right] \Bigg\} \sim \sum_n \frac{\widetilde R_n}{\wn- \wn_n} ~.
\end{equation}
The poles and its residues can be extracted from the following representation of the digamma function
\begin{equation}\label{eq:digammaexp}
\psi(x) = -\gamma_{\rm E} -\sum_{n=1}^\infty \left( \frac{1}{x-1+n} -\frac{1}{n} \right) ~,
\end{equation}
so the residues will scale as $\widetilde{R}_n \sim \wn^2 \,r_n$, with $r_n\sim 1$ and the frequencies $\wn_n\sim n$. Evaluated at the pole this gives $\widetilde{R}_n\sim \wn_n^2 \,r_n\sim n^2$, as was confirmed numerically in \cite{Amado:2007yr}. Using current conservation, the density-density correlator is related to the longitudinal current-current correlator through $G_{tt} = \frac{q^2}{ \omega^2} G_{ii}^L$, so the residues of the density-density correlator go as $R_n\sim q^2 \,r_n \sim q^2$. 

The imaginary part of the retarded correlator receives a contribution from the quasinormal modes
\begin{equation}
\Im \widetilde{G}_{\rm R} \sim \omega \sum_n \frac{\Im \left(\omega_n^2 R_n\right)}{ \left(\omega^2-\Re\omega_n^2\right)^2+\left(\Im \omega_n^2\right)^2} \sim q^2 \omega \sum_n \frac{1}{n^2} ~.
\end{equation}
The sum over the quasinormal modes is convergent in this case. On the other hand the real part goes as
\begin{equation}
\Re \widetilde{G}_{\rm R} \sim \sum_n \frac{|\omega_n|^2 \Re \left(\omega_n^* R_n\right)}{ \left(\omega^2-\Re\omega_n^2\right)^2+\left(\Im \omega_n^2\right)^2} \sim q^2 \sum_n \frac{1}{n} ~.
\end{equation}
In this case the sum is not convergent, although the Green function is finite. The divergence of the sum over quasinormal modes can be cured by subtracting a divergent analytic part, or in a better defined way, subtracting order by order a small analytic part. The analytic term would look
\begin{equation}
\cA =\sum_n \cA_n \sim -q^2 \sum_n \frac{1}{n} ~.
\end{equation}
In order to give a consistent approximation to the retarded correlator, the analytic term must be properly taken into account. For the first $m$ modes
\begin{equation}\label{eq:Grapprox}
\widetilde{G}_{\rm R}(\omega, q) \simeq \frac{-i q^2 \sigma_{\rm H}(q)}{\omega +i q^2 D_{\rm H}(q)} +q^2 \sum_{n=1}^m\left( \frac{r_n(q)}{\omega-\omega_n(q)} + \frac{-r_n^*(q)}{\omega+\omega_n^*(q)}\right)-C_m q^2, \ \ \omega, q < T ~,
\end{equation}
where we have defined the hydrodynamic pole and residue in terms of a $q$-dependent `conductivity' and `diffusion coefficient' $R_{\rm H}=-i\,q^2 \sigma_{\rm H}(q)$ and $\Omega_{\rm H}=-i\,q^2 D_{\rm H}(q)$. In the zero momentum limit, they take the value of the known constant transport coefficients $D=1/(2 \pi T)$, $\sigma=N^2 T/16 \pi$. The coefficient of the analytic term $C_m$ depends on the number of modes and diverges when $m\to \infty$. The exact Green function including all the modes would differ by a small amount from this approximation at small enough frequencies.

In the hydrodynamic approximation only the mode associated to diffusion would be considered, adding higher derivative corrections to the effective description as momentum increases. It is then interesting to find out its exact behaviour beyond the hydrodynamic limit from the holographic dual. In ref.\cite{Amado:2007yr}, the quantities $\sigma_{\rm H}(q)$ and $D_{\rm H}(q)$ were computed numerically. It was found that the residue of the diffusion mode is oscillatory, so for a finite value of $q< 2\pi T$, the $q$-dependent conductivity we have defined becomes zero and then negative. Notice that the analytic corrections that can come from higher modes would not solve this problem, so clearly the diffusion mode does not have a good hydrodynamical interpretation anymore. Furthermore, the residue vanishes exponentially with increasing momentum, so this mode decouples and the system leaves the hydrodynamic regime. In the section \ref{sec:metric} we will find an analogous behaviour in the shear diffusion mode of the stress tensor two point function and explain its properties there in more detail.

\section{\label{sec:vector}More on R-charge diffusion}
We now consider the conserved current $J_\mu$ associated to the global R-symmetry in the $\Nfour$ theory
\begin{equation}
\partial^\mu J_\mu=0 ~.
\end{equation}
The diffusion of the R-charge $Q=\int \dd^3\bx \,J_0$ is described by a hydrodynamic mode, that at low frequencies and momenta dominates the response of the system to small perturbations $j_\mu$
\begin{equation}
\vev{\delta J_\mu(t,\bx)} =-\int \dd t'\,\dd^3\bx' \,G_{\mu\alpha}(t-t', \bx-\bx ') \,j^\alpha(t',\bx ') ~.
\end{equation}
The global current $J_\mu$ is dual to a five-dimensional gauge field $A_M$.\footnote{That is actually a component of the metric with the group index associated to an internal space.} We work in the gauge $A_r=0$. The gauge invariant quantities are the longitudinal and transverse electric fields $E_L = q A_0 + \omega\,\bq\cdot\mathbf{A}/q$, $\mathbf{E}_T = \omega \mathbf{A}_T$, where $\bq \cdot \mathbf{A}_T=0$. We describe the method to find the quasinormal modes in the appendix \ref{sec:method}.

The retarded two-point functions can be expressed in terms of two scalar functions $\Pi_T$ and $\Pi_L$ corresponding to transverse and longitudinal polarizations, assuming $\bq =(0,0,q)$
\begin{align}
\widetilde{G}_{TT}& = \Pi_T ~,& \widetilde{G}_{tt}& = \frac{q^2}{\omega^2 - q^2} \,\Pi_L ~, \\
\widetilde{G}_{tz}(\omega,q)& = \frac{\omega}{q} \,\widetilde{G}_{tt}(\omega,q) ~,& \widetilde{G}_{zz}(\omega,q)& =\frac{\omega^2}{q^2} \,\widetilde{G}_{tt}(\omega,q) ~.
\end{align}
The quasinormal modes and residues $R^{T,L}_n$ of the scalar functions $\Pi_{T,L}$ were computed in \cite{Amado:2007yr}. We remark that the residues of the density-density correlator are actually
\begin{equation}
R_n^{(tt)}= \frac{q^2}{ \omega_n^2(q)-q^2} \,R_n^{(L)}(q) ~.
\end{equation}
In \cite{Amado:2007yr}, we found that the residue of the diffusion mode shows a damped oscillatory behaviour with the momentum. The zeroes appear when the diffusion pole reaches the value of the imaginary part of the quasinormal frequencies at zero momentum. The values of the momentum when this happens can be found analytically (appendix \ref{sec:zeroes}). At a slightly smaller value of the momentum the imaginary part of the $n^{\rm th}$ higher quasinormal mode becomes smaller than the value for the hydrodynamic mode, i.e. the diffusion mode crosses the $n^{\rm th}$ quasinormal mode. The time development is then dominated by the contribution of the lowest quasinormal modes instead of the hydrodynamic mode. 

A way to estimate the regime of validity of the hydrodynamic approximation is to introduce an external perturbation, localized in time or space and compute the hydrodynamic time (\ref{eq:hydro_time}) or length (\ref{eq:hydro_length2}) scales. In the first case we consider the evolution of the charge density (given by $G_{tt}$), while in the second case we do it for the longitudinal component of the current (given by $G_{zz}$). The time-localized source is a plane wave of fixed momentum $q$ that begins at $t=0$ and last some time $\Delta t$. The perturbation creates inhomogeneities that start to relax to equilibrium after it is switched off. The late time relaxation will be dominated by the diffusion mode. If the perturbation lasts for a long time, it is possible for the system to reach a steady state already dominated by the diffusion mode. This seems to be the case as long as $\Delta t$ is a few times the inverse temperature. If the perturbation is short lived, then we find that the minimal time to enter the diffusion dominated regime is  $2 \pi T \tau_{\rm H}\sim 0.35$. We also find that for short wavelengths, when $q$ approaches the value of the zero of the hydro residue, $\tau_{\rm H}$ grows unbounded. So for wavelengths of the order of the inverse temperature or higher, the relaxation to equilibrium is never dominated by the diffusion mode, but by higher modes.

The space-localized source is a pulse constant on $x$ and $y$ directions and localized in the $z$ direction with size $\Delta z$. The pulse is oscillating with some fixed frequency $\omega$. The perturbation will be screened as we move in the $z$ direction far from the plane where the pulse is localized. At large enough distances, the diffusion complex momentum mode would dominate the decay of the perturbation. We find that if the size of the pulse is larger than the scale given by the temperature $\Delta z > 1/T$ and the frequency is low enough $\omega < 2 \pi T $, the screening is dominated by the hydro mode at any distance. However, at higher frequencies the diffusion mode starts dominating at larger distances from the source, so $\ell_{\rm H}>0$. When the size of the source is of the order of the inverse temperature, finite size effects like the shape of the perturbation start to be important to determine the hydrodynamic length scale. In our definition of hydrodynamic time and length scales there is an explicit dependence on the source, 
\begin{equation}
\tilde\jmath = \left\{\begin{array}{cc} \frac{\sin(\omega \Delta t)}{\omega}  & {\rm localized\ in\ time} \\ & \\ \frac{\sin(q \Delta x)}{q}  & {\rm localized\ in\ space}
\end{array} \right.
\end{equation}
The source is evaluated at the value of the of the mode $\omega_n(q)$ or $q_n(\omega)$. For large values of the frequency or the momentum, the $\sim 1/\omega , \, 1/q$ scaling determines the response, while for very low values it is constant. For intermediate values however, the sine function introduces an oscillatory behaviour. In general, we will find different answers for different kind of perturbations, depending on their shape. 

In summary, no matter the value of $q$ for the time-dependent source there is a minimal hydrodynamic time scale $\tau_{\rm H}$ if the duration of the perturbation is short enough. However, for the space-localized source one can always lower the frequency to a value where the length $\ell_{\rm H}$ becomes zero, no matter how small the source is.

The validity of the hydrodynamic regime can also be studied through the spectral function. As a first approximation we could pick only the diffusion mode. However, the spectral function and the residue have the same sign, so the approximation would clearly fail at the first zero. This means that in order to have a sensible approximation for larger values of the momentum, we must take into account higher modes. We will see that this solves the problem.

\begin{figure}[!htbp]
\centering
\includegraphics[width=7cm]{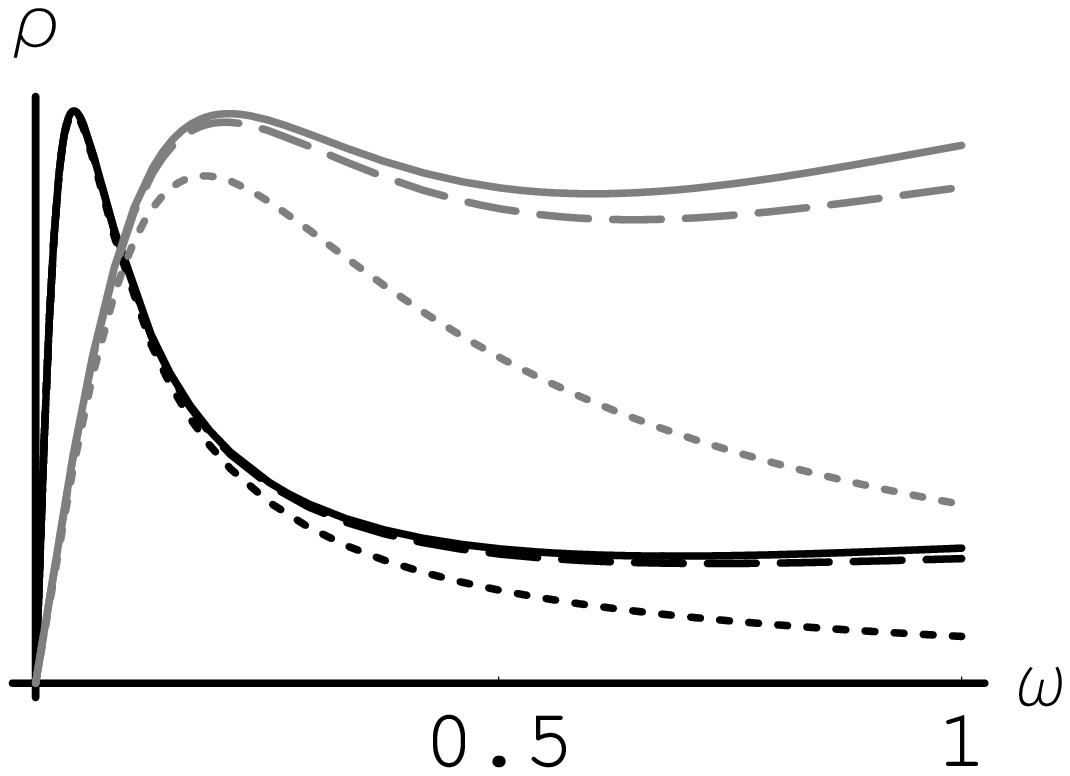} \hfill \includegraphics[width=7cm]{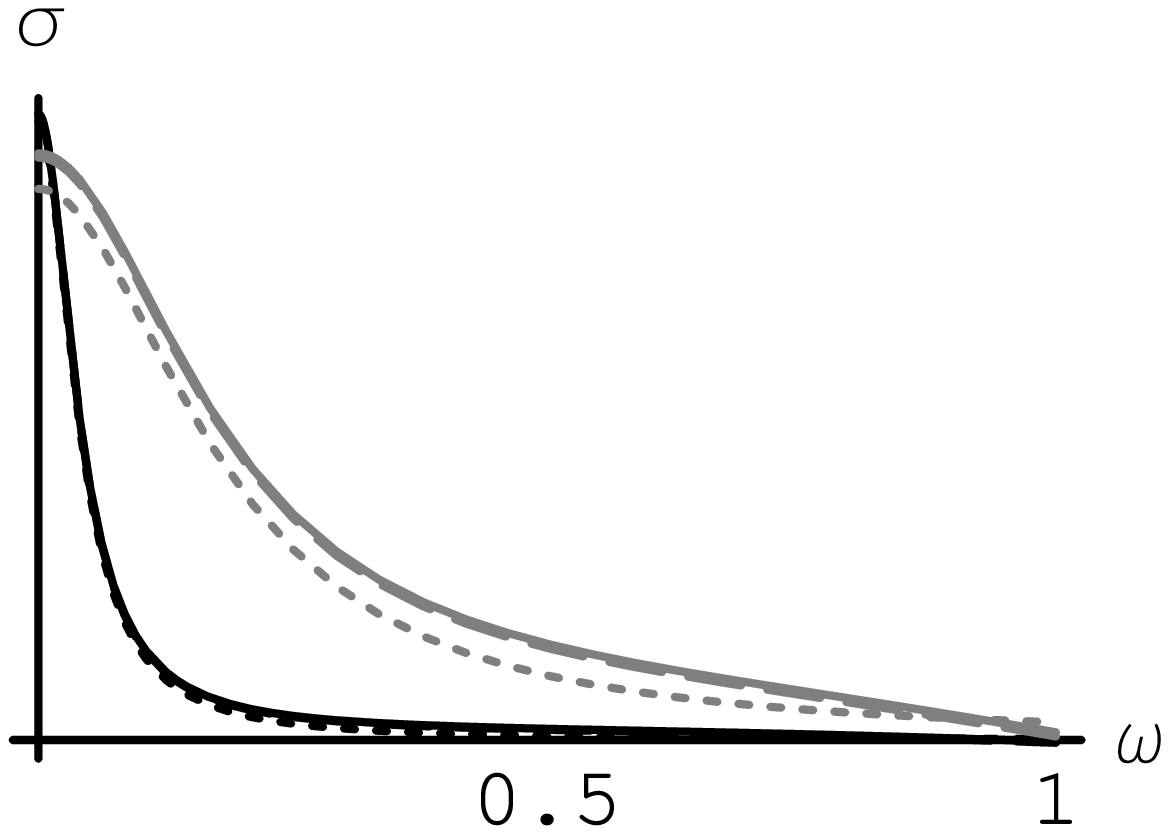} \\
\caption{\label{fig:diffw} Imaginary (left) and real (right) parts of the retarded  $G_{tt}$ correlator as a function of the frequency at $\qn=0.2$ (black) and $\qn=0.4$ (grey). The dotted line is the hydrodynamic mode contribution, the solid line is the exact solution and the dashed line is the four-mode approximation.}
\end{figure}

Following the discussion in section \ref{sec:resonances}, we can compare the results from the exact retarded density-density correlator $\widetilde{G}_{tt}$ with the approximation taking the first four quasinormal modes or just the hydrodynamic mode. In order to fix the analytic term $C_4$ in (\ref{eq:Grapprox}), we equate the values of the exact and the approximate Green functions at zero frequency and $\qn=0.2$, giving $C_4 \simeq 3.9$. In figure \ref{fig:diffw} we can see that the approximation for fixed momentum is very good in the interval $\wn < 1$, even for larger values of the momentum. 

\begin{figure}[!htbp]
\includegraphics[scale=0.9]{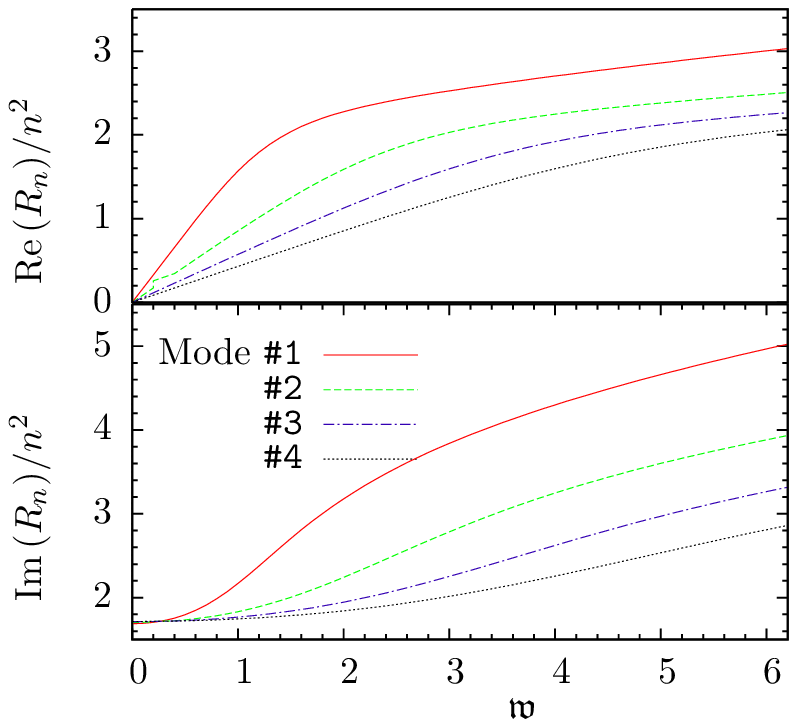}\hfill
\includegraphics[scale=0.9]{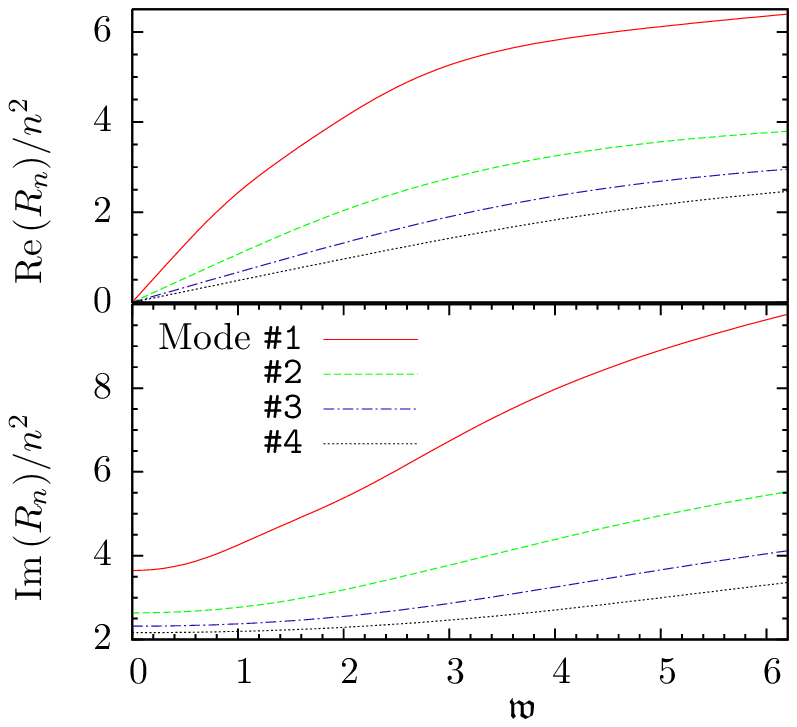} 
\caption{\label{fig:ResCMMvec}(Left) Real and imaginary parts of the residues for the first four complex momentum modes in the transverse component $\Pi_T$. (Right) Idem for longitudinal component $\Pi_L$. The numerical values are normalized by $N^2 T^2/8$ and the square of the mode number.}
\end{figure}

\begin{figure}[!htbp]
\centering
\includegraphics[scale=0.9]{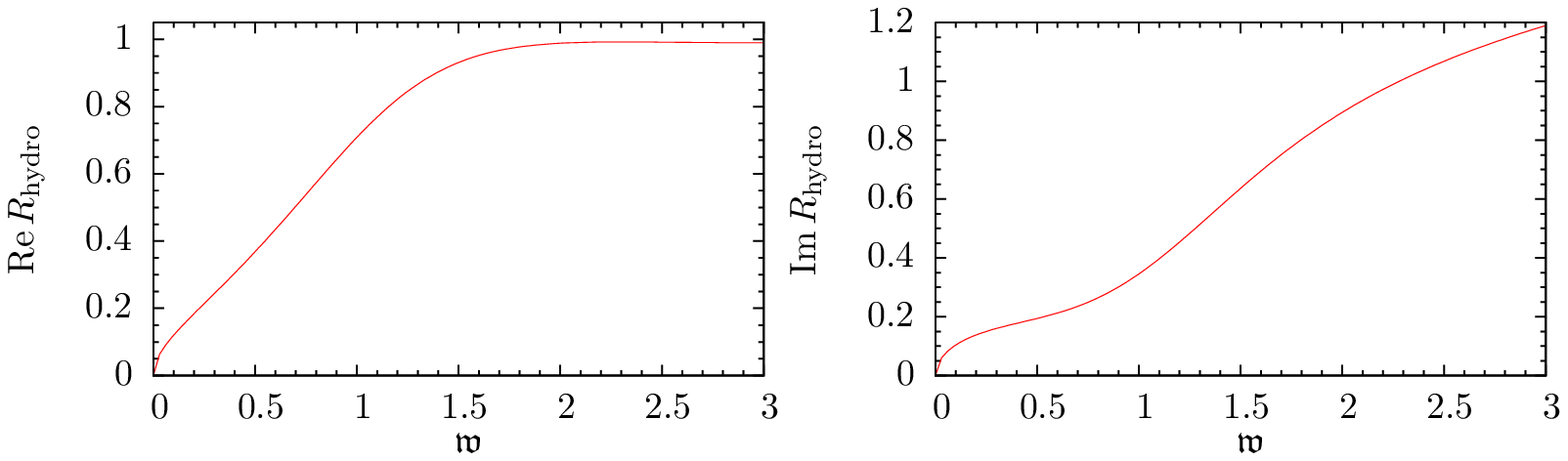}
\caption{\label{fig:ResCMMHvec}Real and imaginary parts of the residues for the diffusion complex momentum mode. The numerical values are again expressed in terms of $N^2 T^2 /8$.}
\end{figure}

We also present the residues for the complex momentum modes, in figures \ref{fig:ResCMMvec} and \ref{fig:ResCMMHvec}. Again these are the residues of the scalar functions $\Pi_{T,L}$. We find that contrary to the quasinormal modes, the hydrodynamic mode does not decay at high frequencies (see figure \ref{fig:ResCMMHvec}).  In \cite{Amado:2007pv} the locations of the complex momentum modes in the R-charge diffusion channel have been studied. It was found numerically that the real  part of the complex momentum approaches $\omega$ whereas the imaginary part becomes smaller and smaller at high frequency for all the modes. Furthermore we argued in the introduction that the high frequency and momentum limit can be exchanged with the zero temperature limit $T\rightarrow 0$. The theory at $T=0$ recovers Lorentz symmetry and in the case at hand even conformal symmetry. This restricts all signal propagation at $T=0$ automatically to the light cone, showing that indeed $v_F=1$.


\section{\label{sec:metric}Energy-momentum tensor correlators}
The energy-momentum tensor is a conserved quantity
\begin{equation}
\partial^\mu T_{\mu\nu}=0 ~,
\end{equation}
we can identify $ E=\int \dd^3\bx \,T_{00}$ and $P_i=\int\dd^3\bx \,T_{0 i}$ as the conserved energy and momentum. Since they are conserved and cannot be dissipated away, they will slowly spread through the plasma or they will be displaced between different regions. This is described by the hydrodynamic shear and sound modes. The response to an external perturbation represented by the source $j_{\mu\nu}(t,\bx)$ is given by
\begin{equation}
\vev{\delta T_{\mu\nu}(t,\bx)} =-\int \dd t'\,\dd^3\bx' \,G_{\mu\nu,\,\alpha\beta}(t-t', \bx-\bx ') \,j^{\alpha\beta}(t',\bx ') ~,
\end{equation}
where $G_{\rm R}$ is the retarded two-point function. For low frequencies and momentum (large times and distances), the response is dominated by the hydrodynamic modes.

The energy-momentum tensor $T_{\mu\nu}$ in the gauge theory is dual to the five-dimensional metric $g_{MN}$. We work in the linear approximation, so we will consider small fluctuations $h_{MN}$ on top of the background metric of a planar black hole in AdS${}_5$ space \erf{eq:AdSBH}. We fix the gauge to $h_{rM}=0$ and expand the metric components in plane waves $h_{\mu\nu}(t,\bx,r)=\e^{-i\omega t+i\bq \bx} h_{\mu\nu}(r)$. It is convenient to use quantities that are gauge-invariant under the residual diffeomorphism transformations and group them according to their representation under the rotational group. In this way, we can distinguish three different components, scalar, vector and tensor, that correspond to the sound, shear and scalar channels in the dual theory \cite{Kovtun:2005ev}. Assuming that the momentum is along the $z$ direction, and using the dimensionless frequency and momentum $2\pi T (\wn,\qn) :=(\omega,\bq)$, the gauge-invariant quantities are
\begin{eqnarray}\label{eq:metriccomp}
\notag  ({\rm Shear})\ \ {Z_1}_i & = & \qn h_{ti}/r^2+ \wn h_{zi}/r^2, \quad i=x,y ~, \\
\notag  ({\rm Sound})\ \ Z_2 & = & \qn^2 h_{tt}/r^2+2 \wn \qn h_{tz}/r^2+ \wn^2 h_{zz}/r^2+(\qn^2-\wn^2+ r f'(r)/2) (h_{xx}+h_{yy})/r^2 ~, \\
({\rm Scalar})\ \ Z_3 & = & h_{xy}/r^2 ~. 
\end{eqnarray}
The retarded Green function of the dual gauge theory is computed as the boundary action of classical solutions satisfying infalling boundary conditions at the horizon, taking into account the subtleties of the Lorentzian formulation \cite{Son:2002sd, Herzog:2002pc}. The poles of the retarded correlator correspond to the quasinormal modes, that are normalizable solutions at the AdS boundary. The method to find the quasinormal modes is explained in appendix \ref{sec:method}. The Green functions $G_1, G_2, G_3$ found using the $Z_1, Z_2, Z_3$ components are scalar quantities. They are the coefficients of the tensor projectors depending on $\omega$ and $q$ in which energy-momentum correlators can be decomposed once Lorentz invariance is reduced to rotational invariance (c.f.~\cite{Kovtun:2005ev}). For instance, 
\begin{eqnarray}\label{eq:tensorgreen}
G_{tx,\,tx} & = & \frac{1}{2} \frac{q^2}{\omega^2-q^2} \,G_1 ~, \\
G_{tt,\,tt} & = & \frac{2}{3} \frac{q^4}{(\omega^2-q^2)^2} \,G_2 ~, \\
G_{xy,\,xy} & = & \frac{1}{2} \,G_3 ~.
\end{eqnarray}
We also compute the residues of the different modes. Notice that we are using the functions $G_1$, $G_2$ and $G_3$, so in order to recover the right residue for the different components of the energy-momentum tensor correlators, one should take into account the appropriate factors coming from the tensor projectors.

\begin{figure}[!htbp]
\centering
\includegraphics[width=4.5cm]{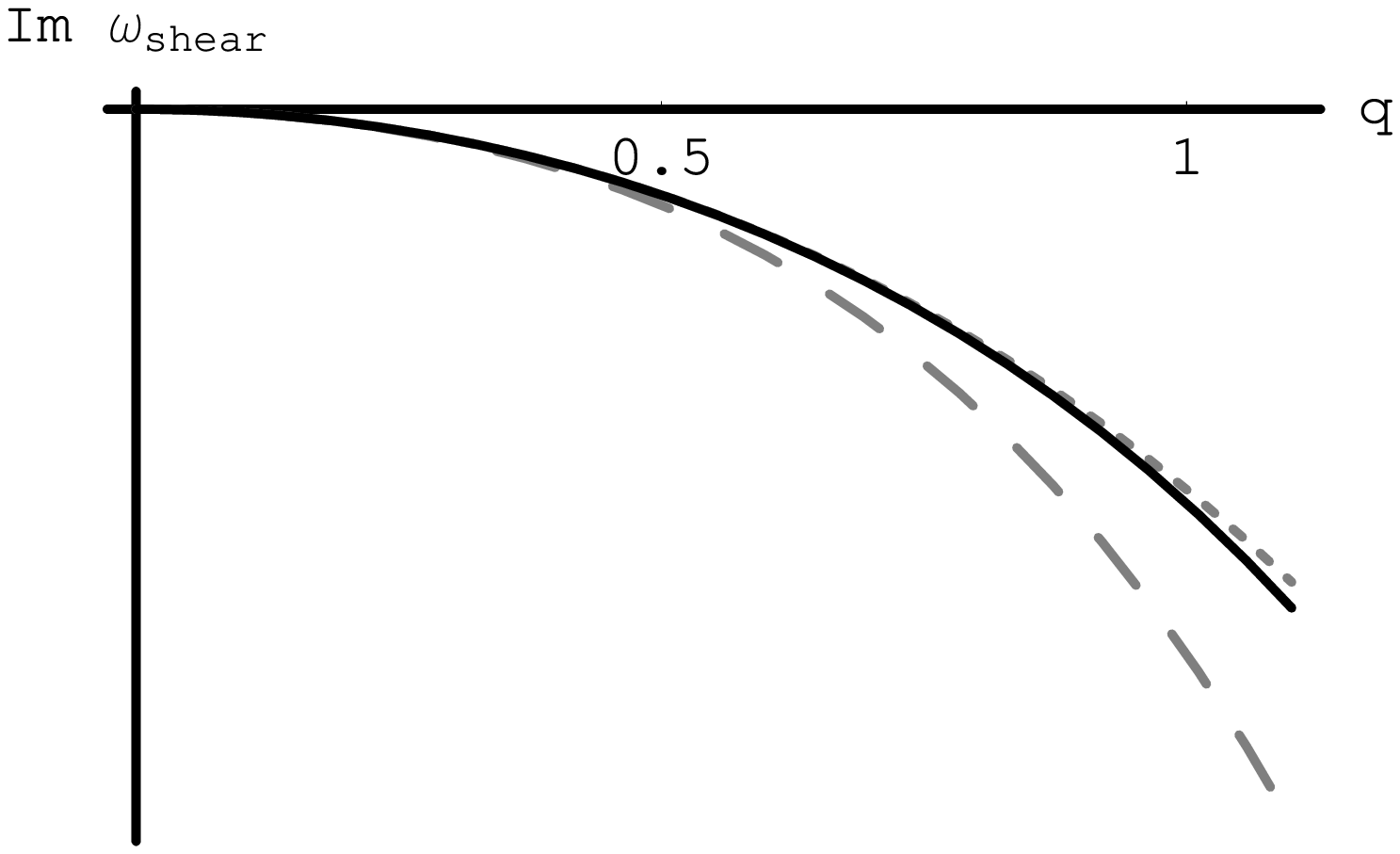} \hfill
\includegraphics[width=4.5cm]{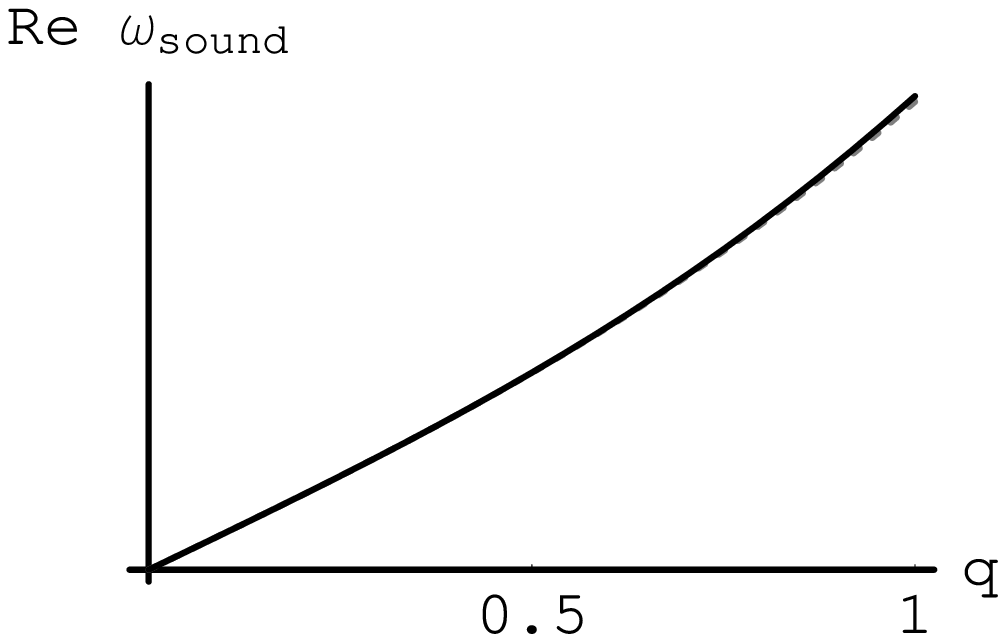} \hfill
\includegraphics[width=4.5cm]{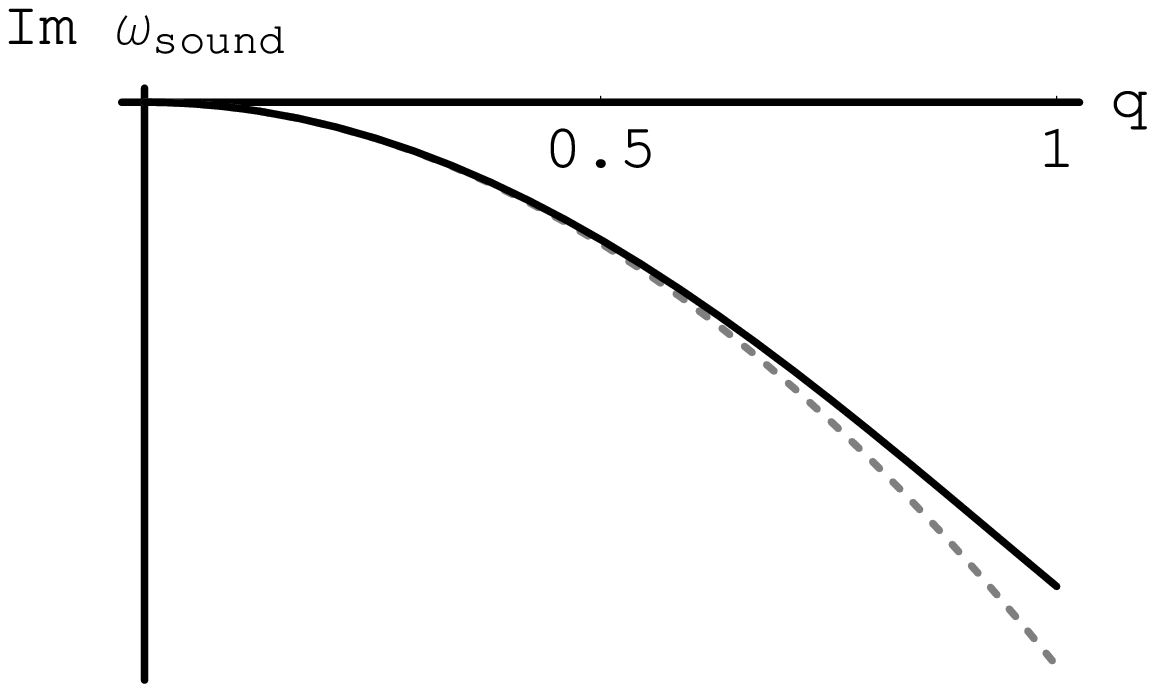} 
\caption{\label{fig:hydromodes}(Left) The numerical value of the shear mode [solid black] compared with the correct expression [dotted gray] and the value from second order hydrodynamics [dashed gray]. (Middle) The numerical value of the real part of the sound mode [solid black] compared with the second order hydrodynamic approximation [dotted gray]. (Right) The numerical value of the imaginary part of the sound mode [solid black] compared with the second order hydrodynamic approximation [dotted gray].}
\end{figure}

In the vector and scalar channels, the lowest quasinormal mode shows the right behaviour at low momentum to be identified with the shear ($\omega =-iD q^2$) and sound ($\omega= v_s q- i\Gamma_s q^2 $) modes respectively. In figure \ref{fig:hydromodes} we compare our numerical results with the second order hydrodynamic approximations of \cite{Baier:2007ix}. From the Kubo formula of the tensor component $G_{xy,xy}$ the second order corrections including a term $\eta \,\tau_\Pi \,\omega^2$ have been identified with 
\begin{equation}
\tau_\Pi=\frac{2 -\log 2 }{ 2 \pi T} ~.
\end{equation}
According to the hydrodynamic expansion, the first corrections to the shear and sound poles would be
\begin{eqnarray}
\notag \omega_{\rm shear} & = & -i \frac{\eta}{ s T} q^2-i \left(\frac{\eta}{s T}\right)^2 \tau_\Pi  \,q^4 ~, \\
\omega_{\rm sound} & = & v_s q -i \Gamma_s q^2 + \frac{\Gamma_s}{v_s}\left( v_s^2 \,\tau_\Pi -\frac{\Gamma_s}{2}\right) q^3 ~.
\end{eqnarray}
With $\eta$ the shear viscosity, $s$ the entropy density and $v_s$ and $\Gamma_s$ the sound velocity and the sound attenuation. However, in the holographic computation the dispersion relation for the shear mode differs from the second order hydrodynamic result.  This discrepancy can be attributed to the necessity of going to third order hydrodynamics to capture the right $O(q^4)$ correction to the shear pole. In terms of the dimensionless variables the holographic theory gives the next to leading order corrections in the dispersion relations
\begin{eqnarray}
\notag \wn_{\rm shear} & = & -i \,\frac{\qn^2}{2}-i \frac{(1-\log 2)}{4} \,\qn^4 ~, \\
\wn_{\rm sound} & = &  \frac{\qn}{\sqrt{3}} -i \,\frac{\qn^2}{3} + \frac{3-2 \log 2}{6 \sqrt{3}} \,\qn^3 ~.
\end{eqnarray}
When we compare the numerical results with these (fig.~\ref{fig:hydromodes}), we find a very good agreement up to $\qn \lesssim 1$. Similar comparisons between the exact values and first order hydrodynamics or the first corrections can be found in \cite{Nunez:2003eq, Kovtun:2005ev, Baier:2007ix}. It seems justified then to consider just the first hydrodynamic corrections to describe the properties of the shear and sound modes up to frequencies of the order of the temperature $\omega, q \sim T$. Note however, that the $q^4$ term that arises from second order hydrodynamics obviously does not approximate the numerical result very well, clearly indicating that second order hydrodynamics is not valid to predict the $q^4$ terms as already pointed out in \cite{Baier:2007ix}.

\begin{figure}[!htbp]
\includegraphics[scale=0.9]{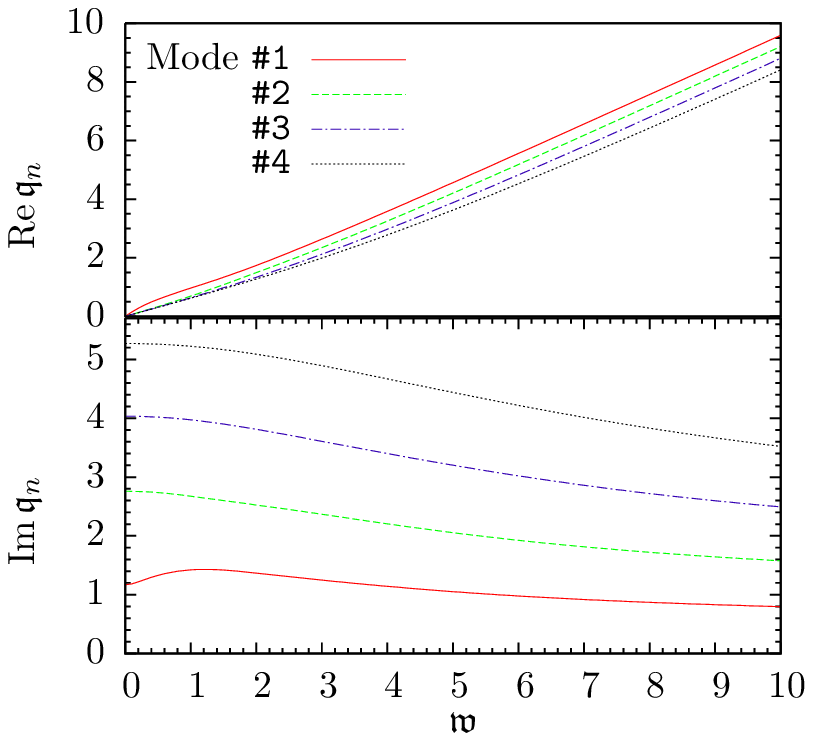}\hfill
\includegraphics[scale=0.9]{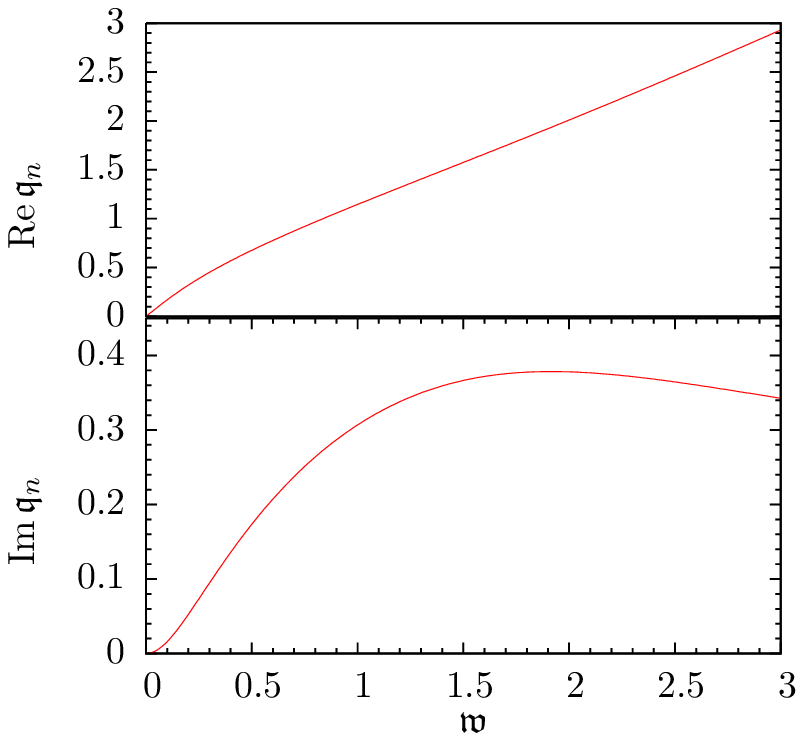} 
\caption{\label{fig:CMMsound}(Left) Real and imaginary parts of the first four complex momentum modes in the sound channel. (Right) Idem for sound mode.}
\end{figure}

\begin{figure}[!htbp]
\includegraphics[scale=0.9]{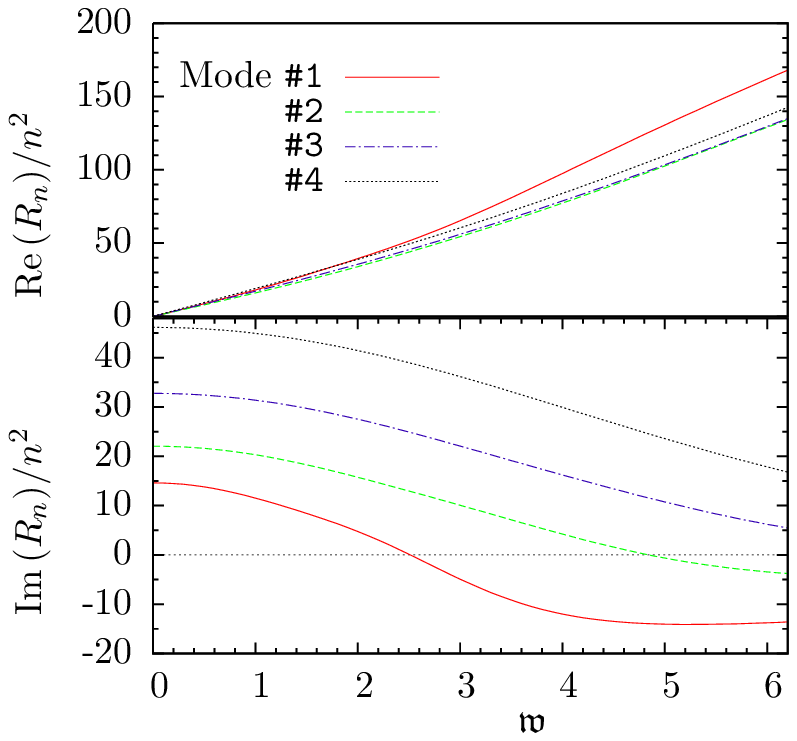}\hfill
\includegraphics[scale=0.9]{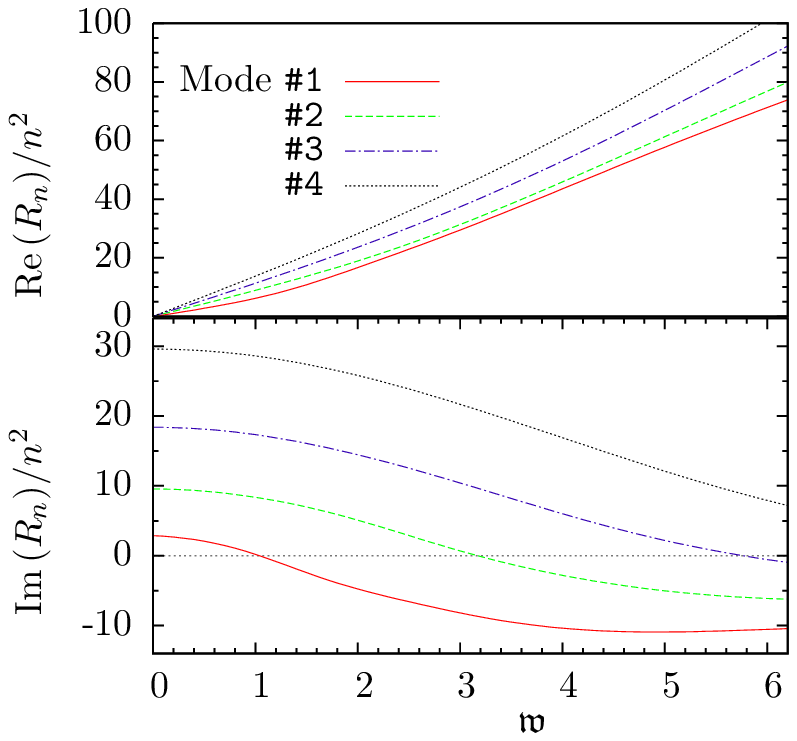} 
\caption{\label{fig:ResCMMmetric}(Left) Real and imaginary parts of the residues for the first four complex momentum modes in the shear channel. (Right) Idem for sound channel.}
\end{figure}

\begin{figure}[!htbp]
\includegraphics[scale=0.9]{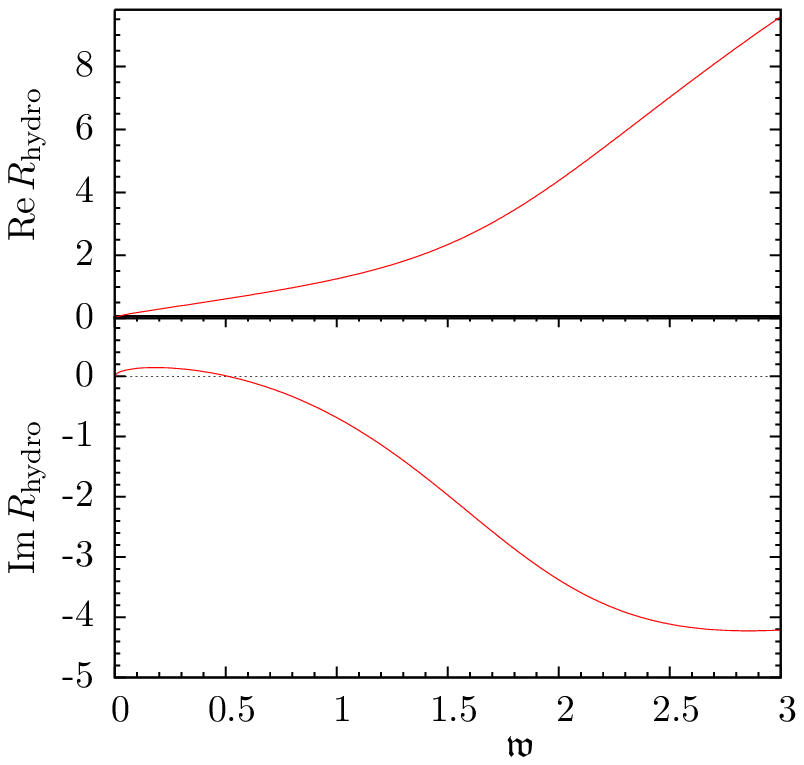}\hfill
\includegraphics[scale=0.9]{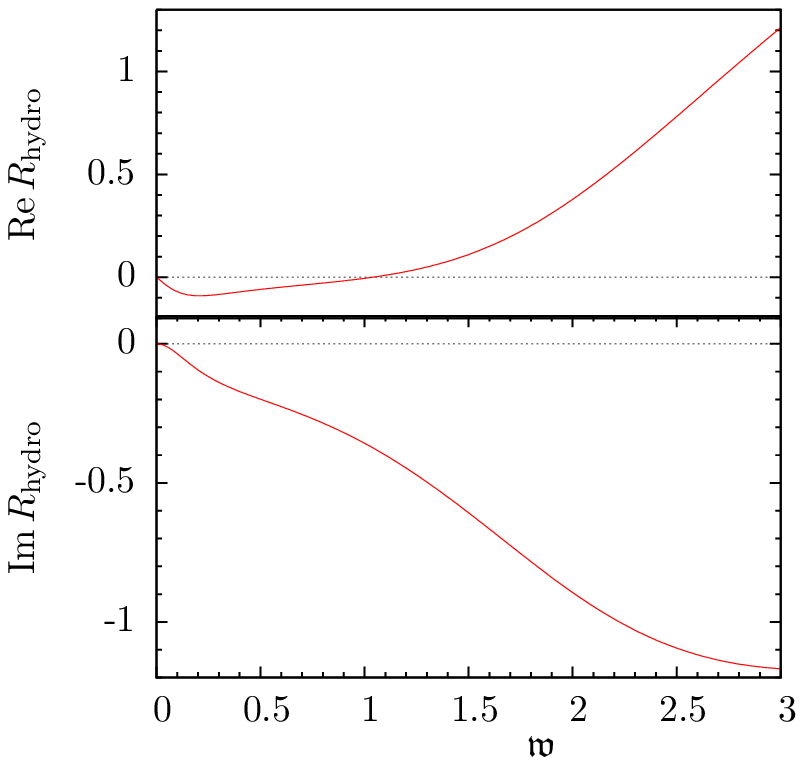} 
\caption{\label{fig:ResCMMHmetric}(Left) Real and imaginary parts of the residue for the shear mode. (Right) Idem for the sound mode.}
\end{figure}

We have also computed the complex momentum modes that describe the penetration depth of a perturbation inside the plasma as a function of the frequency \cite{Amado:2007pv}. Here we show the complex momentum modes for the sound channel in figure \ref{fig:CMMsound}, while for the shear channel can be found in reference above. The results for the residues are displayed in figures \ref{fig:ResCMMmetric} and \ref{fig:ResCMMHmetric}. The zero frequency value corresponds to the inverse of the screening masses in the plasma. Again we find that the imaginary parts become smaller with increasing frequency and that the real parts approach the light cone $\omega/q \to 1$ indicating a front velocity $v_F=1$ as in the other channels (see Appendix \ref{sec:frontveloc}).

\subsection{Residues and higher thermal resonances}
\begin{figure}[!htbp]
\includegraphics[scale=0.9]{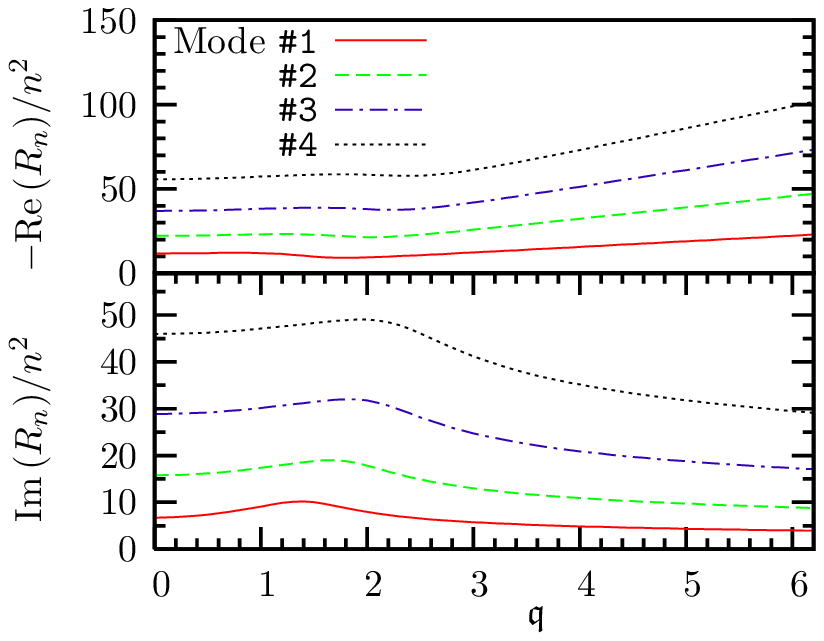}\hfill
\includegraphics[scale=0.9]{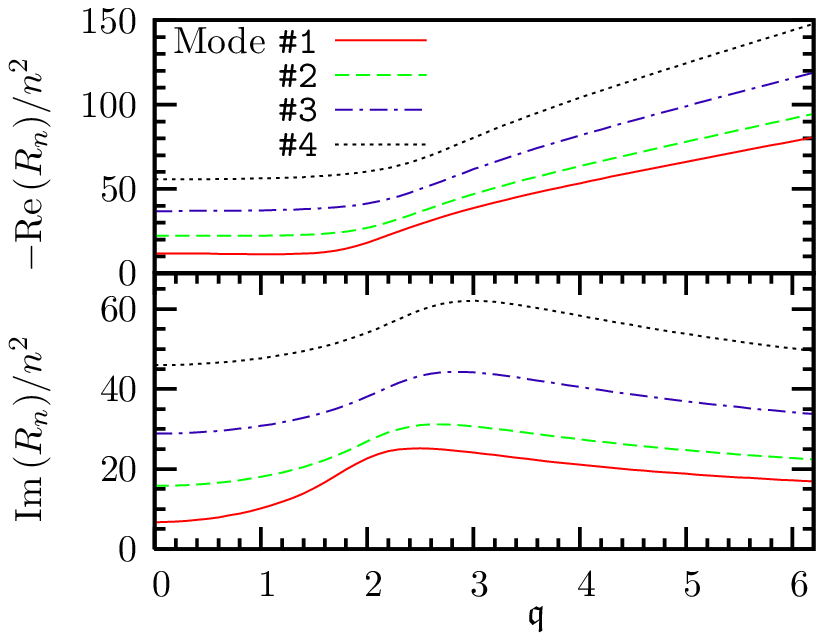} 
\caption{\label{fig:ResQNMmetric}(Left) Real and imaginary parts of the residues for the first four quasinormal modes in the shear channel. (Right) Idem for the sound channel. The numerical values are normalized by $\pi^2 N^2 T^4 $ and the square of the mode number.}
\end{figure}

\begin{figure}[!htbp]
\includegraphics[scale=0.8]{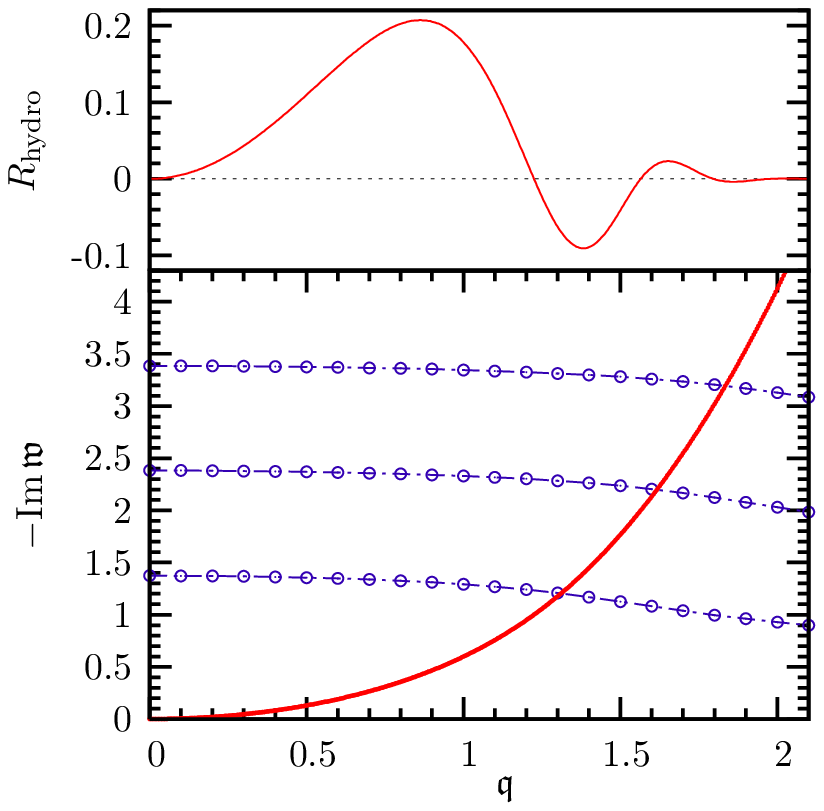}\hfill
\includegraphics[scale=0.9]{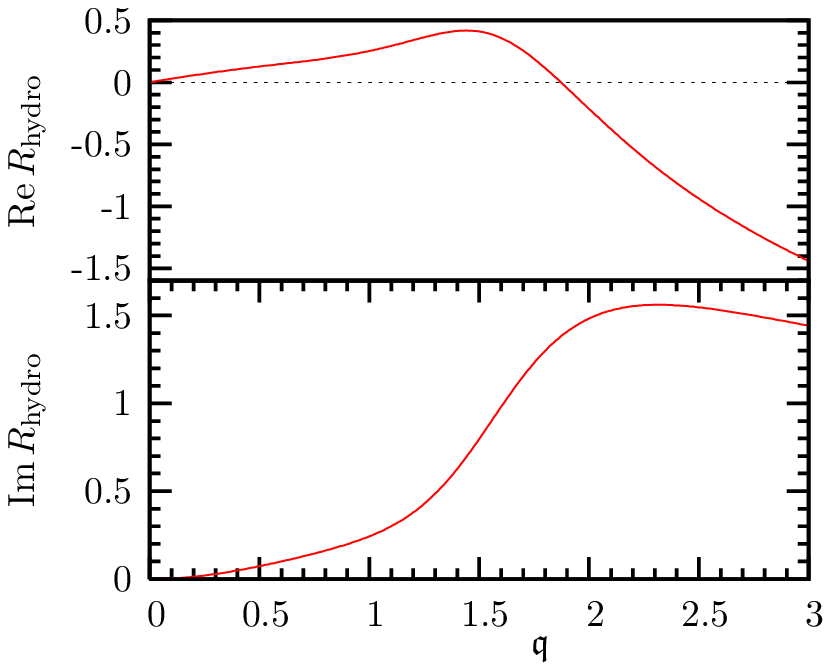} 
\caption{\label{fig:ReHmetric}(Left) When the shear mode hits a special value, there is a qualitative change in the behaviour of the quasinormal modes. Zeroes of the residue of the shear mode roughly coincide with the crossing of the shear mode and the quasinormal modes. (Right) The residue of the sound mode behaves as the rest of the quasinormal modes. The numerical values are normalized by $\pi^2 N^2 T^4$}
\end{figure}

In figures \ref{fig:ResQNMmetric} and \ref{fig:ReHmetric} we plot the residues of the lowest four quasinormal modes in the shear and sound channels. The residue in the shear mode (figure \ref{fig:ReHmetric}) shows a decaying and oscillatory pattern with increasing momentum, similar to the diffusion mode of the R-current. We can apply the same arguments here, at short enough wavelengths the first quasinormal mode will dominate the late time behaviour and the hydrodynamic approximation will not be valid. The sound mode on the other hand has a non-vanishing residue at high momentum, and its pole moves away from the real axis only for low values of the momentum, but then it behaves as an ordinary quasinormal mode, so it always dominates the late time response of the system.

Using the numerical values of the residues, we can study the hydrodynamic time (\ref{eq:hydro_time}) and length (\ref{eq:hydro_length2}) scales for the shear and sound modes. Notice that the residues of the sound channel $R_n^{(2)}(q)$ we have computed  are evaluated at the quasinormal frequencies and correspond to the `scalar' Green function $G_2$. From (\ref{eq:tensorgreen}), we can see that in order to expand $G_{tt,\,tt}$ as a sum over quasinormal modes we should use the residues 
\begin{equation}
R_n^{(tt,\,tt)}=\frac{2}{3} \frac{q^4}{(\omega_n^2(q)-q^2)^2} \,R_n^{(2)}(q) ~.
\end{equation}
In the shear channel we are computing the residues $R_n^{(1)}(q)$ of the function $G_1$ evaluated at the quasinormal frequencies. Using (\ref{eq:tensorgreen}), the residues of $G_{tx,\,tx}$ should be
\begin{equation}\label{eq:resshear1}
R_n^{(tx,\,tx)} = \frac{1}{2} \frac{q^2}{\omega_n^2(q)-q^2} \,R_n^{(1)}(q) ~.
\end{equation}
However, contrary to the cases of the density-density correlator of the R-current or the $G_{tt,\,tt}$ correlator of the sound channel, the large frequency behaviour does not asymptote to a momentum-dependent constant ($\sim q^2$ or $q^4$), but to $\sim q^2 (\omega^2-q^2)$. This means that there should be an implicit $\omega^2$ dependence in the residues, but this is not shown by our computation. A way to partially recover the right dependence is to use the following definition
\begin{equation}\label{eq:resshear2}
R_n^{(tx,\,tx)} = \frac{1}{2} \frac{\omega^2-q^2}{(\omega_n^2(q)-q^2)^2}\, q^2\, R_n^{(1)}(q) ~.
\end{equation}

We consider the same kind of perturbations as for the R-current diffusion mode in section \ref{sec:vector}, one is a spatial plane wave of momentum $q$ that is switched on only during a finite amount of time $\Delta t$. It sources the transverse and longitudinal momentum densities, whose response is given by $G_{tx,tx}$ and $G_{tt,tt}$. The other perturbation is a pulse of size $\Delta z$ localized in the $z$ direction that oscillates with frequency $\omega$. We study the response of the transverse and longitudinal momentum currents given by for $G_{zx,zx}$ and $G_{zz,zz}$. The results for the shear mode are qualitatively the same as for  the diffusion mode. For a short-lived time-localized source there is a minimal hydrodynamic time $2 \pi T \tau_{\rm H}\sim 1.34$. The maximal value of the momentum  beyond which the relaxation to equilibrium is dominated by higher modes at late times is $q \sim 2.6 \pi T$. For a space-localized source, for large enough $\Delta z>1/T$ and small enough frequencies $\omega < 0.6 \pi T $, the screening is well described by the complex momentum shear mode. For smaller sizes or higher frequencies in general the shear mode dominates only at a finite distance from the source $\ell_{\rm H} >0$, unless the frequency is very low.

The sound mode behaves qualitatively different because its residue does not vanish at any value of the momentum. It always dominates the relaxation to equilibrium for long enough times or large enough distances as far as our computation can show. Higher modes dominate the response for a finite time after switching off the source for momenta $q > 1.2 \pi T$. There is no minimal time for the sound mode, but since for large frequencies the sound mode behaves as any other mode, the behaviour will not be hydrodynamic until some finite time has passed if the source is very short-lived. If the source is localized in space, higher modes dominate at a finite distance for values of the frequency $\omega > 2.5 \pi T $. 

We can also study the validity of the hydrodynamic regime through the spectral function. As we found for R-current diffusion, the first zero of the shear residue implies a change of sign for the approximation with the shear mode alone, hence a failure of the hydrodynamic approximation. This happens around $\qn\sim 1.3$. We find that also the spectral function restricted to the sound mode alone changes sign but it is not related to a `crossing'. It coincides approximately with the change of behaviour of the residues that can be observed in figure \ref{fig:ResQNMmetric}, close to $\qn \sim 1.1$. Therefore, in order to have a consistent description of the system for larger values of momentum, it is necessary to take into account higher modes.


\begin{figure}[!htbp]
\centering
\includegraphics[width=7cm]{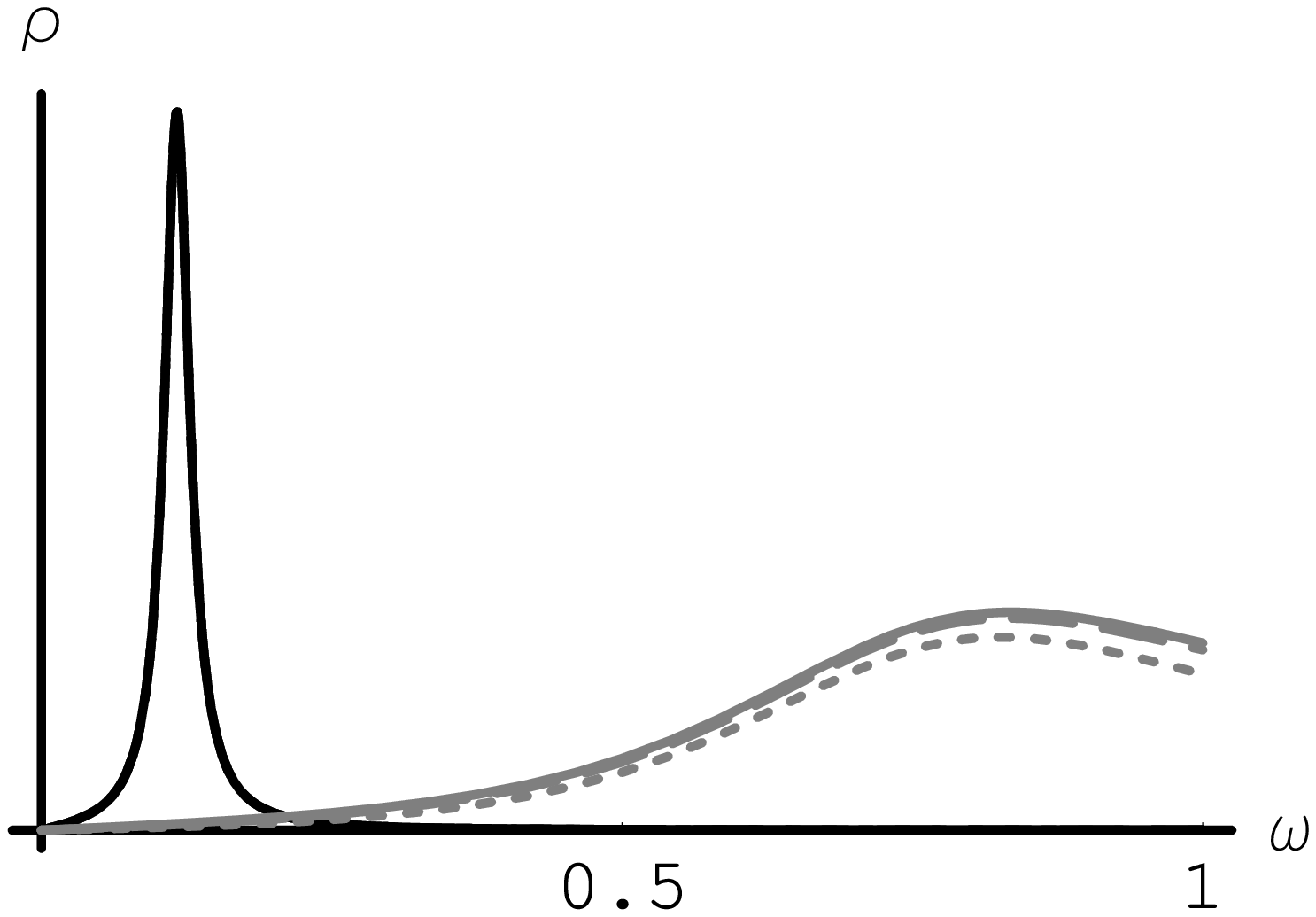} \hfill \includegraphics[width=7cm]{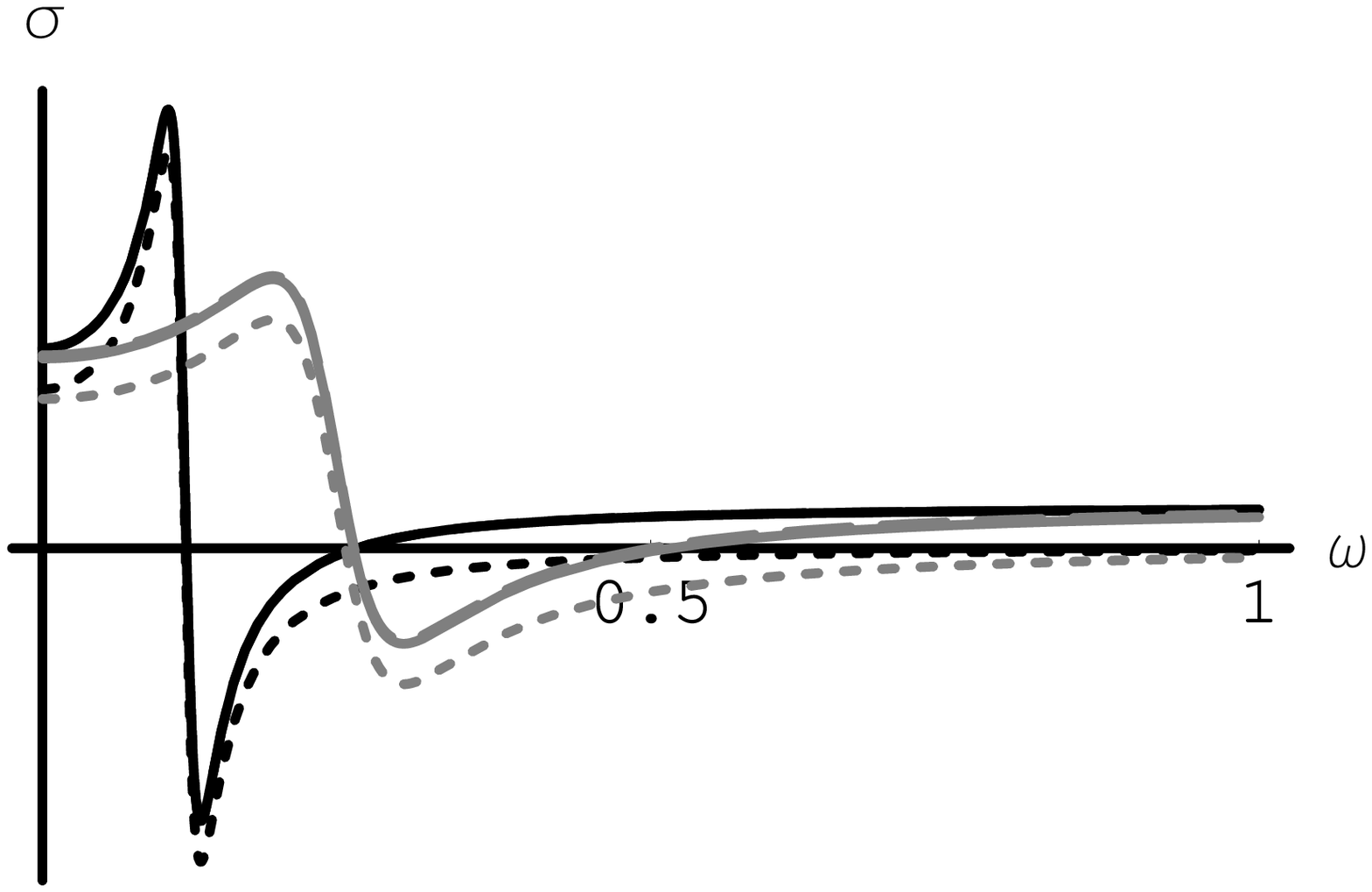} 
\caption{\label{fig:soundw} Imaginary (left) and real (right) parts of the retarded $G_{tt,\,tt}$ correlator as a function of the frequency at $\qn=0.2$ (black) and $\qn=0.4$ (grey) for the real part and $\qn=1$ (grey) for the imaginary part. The dotted line is the hydrodynamic mode contribution, the solid line is the exact solution and the dashed line is the four-mode approximation.}
\end{figure}

Using these results for the residues and the frequencies of the first quasinormal modes we can build an approximation to the retarded correlator as explained in section \ref{sec:resonances}. For the sound channel, the properly defined Green function actually includes the boundary terms found in \cite{Kovtun:2005ev} added to the Frobenius series approximation. Otherwise, an unphysical singularity would appear at $\omega=\pm q$. We examine the $G_{tt,\,tt}$ Green function and find that the approximation and the exact function differ by a real constant $C \sim 0.374$. Once this term is included, they agree for a fair interval of frequencies and momenta (fig.\,\ref{fig:soundw}). Even for values of $\omega$, $q$ smaller but comparable to the temperature, the hydrodynamic mode alone gives a remarkable good approximation for the spectral function. We can see that for $\qn =0.2$, the hydrodynamic approximation and the exact result are virtually indistinguishable for $\wn \lesssim 1$.

Other correlators related to the sound channel can be found from $G_{tt,\,tt}$ using the explicit expressions for the tensor projectors \cite{Kovtun:2005ev}.

\begin{figure}[!htbp]
\centering
\includegraphics[width=7cm]{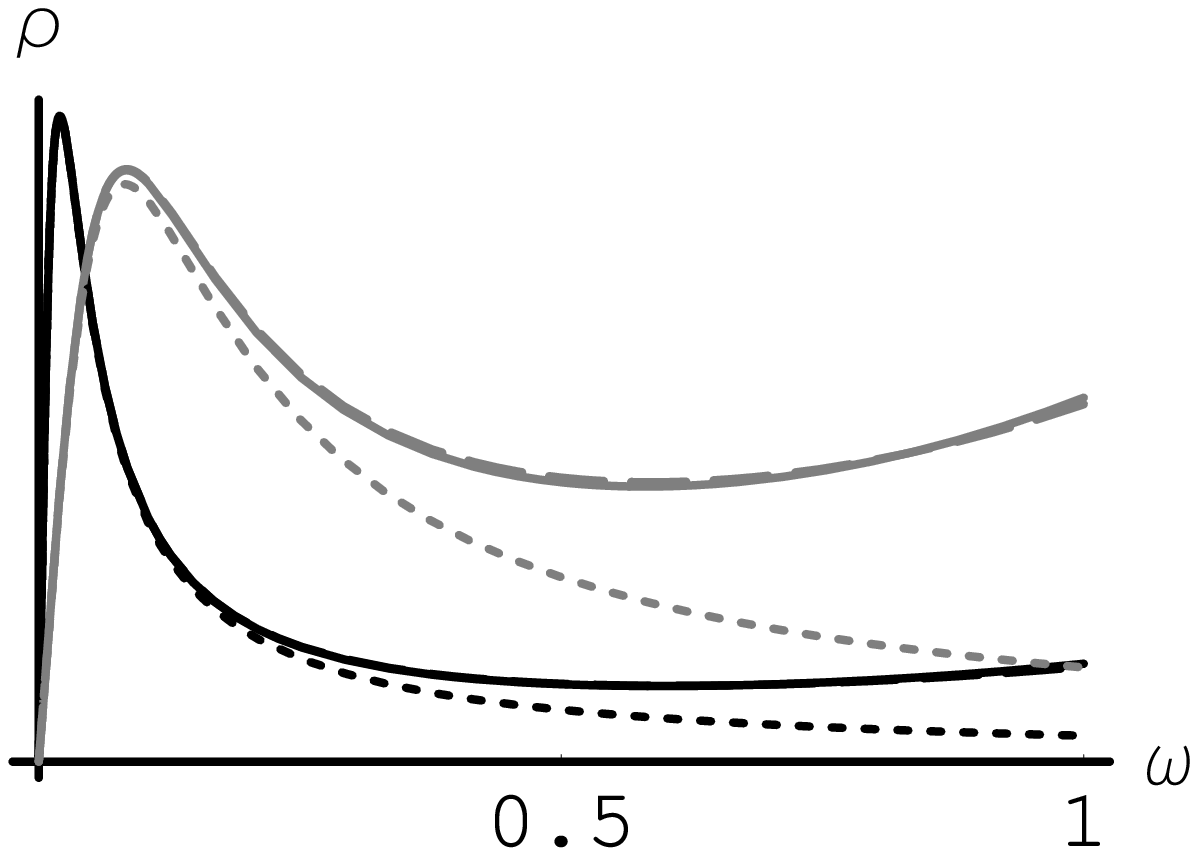} \hfill \includegraphics[width=7cm]{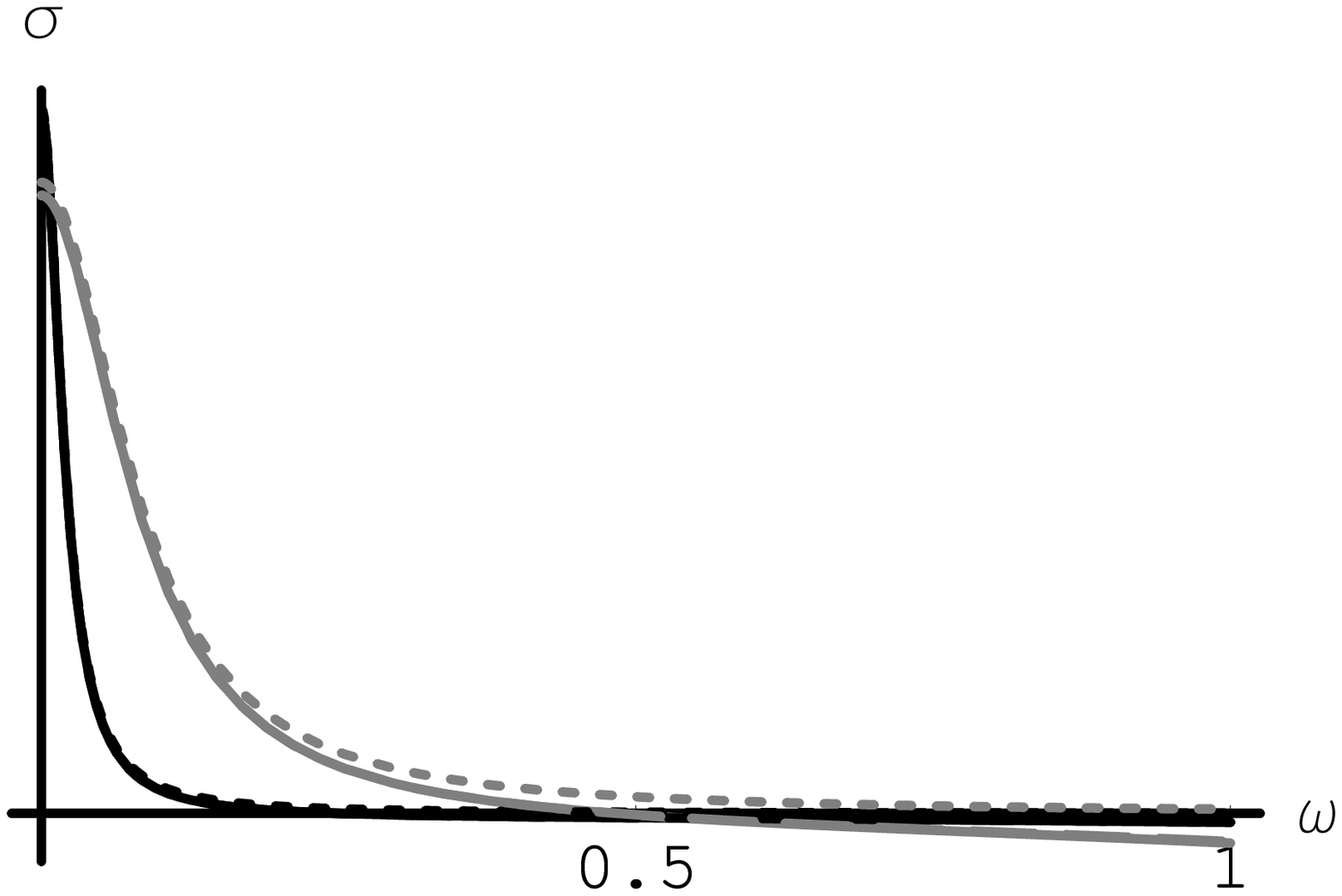} 
\caption{\label{fig:shearw} Imaginary (left) and real (right) parts of the retarded  $G_{tx,\,tx}$ correlator as a function of the frequency at $\qn=0.2$ (black) and $\qn=0.4$ (grey). The dotted line is the hydrodynamic mode contribution, the solid line is the exact solution and the dashed line is the four-mode approximation.}
\end{figure}

We also examine the $G_{tx,\,tx}$ component of the shear channel, the results are in fig. \ref{fig:shearw}. We have to add an analytic piece $\sim 0.9 \qn^2 \wn$ to the spectral function. To the real part of the correlator we need to add a more complicated term $\sim 1.89 \qn^4-0.16 \qn^2-1.77 \qn^2 \wn^2$. The plots made in figure \ref{fig:shearw} have been done using the definition (\ref{eq:resshear2}), we observe very good agreement for $\wn <1,\qn\lesssim 0.6 $. 

\section{\label{sec:results}Discussion}
We have studied hydrodynamics in the strongly coupled $\Nfour$ gauge theory based on linear response theory using the AdS/CFT correspondence. As emphasized we understand hydrodynamics here as the effective theory resulting from integrating out the higher quasinormal modes and keeping only the contribution of the hydrodynamic modes, i.e. those modes whose quasinormal frequency vanishes at zero momentum. We also recall that we can consider this to be a definition of the system being in local thermal equilibrium. One of the important findings is that the hydrodynamic approximation defined in that way has its breakdown built into it: we saw that the diffusion mode in the shear channel crosses the lowest quasinormal mode at around $\qn\approx 1.3$. From that value on it is the lowest non hydrodynamic mode that determines the late time behaviour. In fact, already slightly before the spectral function of the hydrodynamic mode ceases to be positive. We interpret this as a signal for the breakdown of the effective theory based on the hydrodynamic mode alone. Similarly, in the sound channel we found that the spectral function of the sound mode switches sign at $\qn\approx 1.1$ and again we take this as a breakdown of a putative effective theory based on the sound channel alone. The full spectral functions (or even the approximation keeping the contributions from only a few quasinormal modes) behave perfectly reasonable at these points.

An important role in our investigations have been played by the residues. We saw that the shear diffusion mode and the R-charge diffusion are very similar. Both decay in an oscillatory pattern with decreasing frequency and numerically we find that they decouple for momenta $\qn > 1$. As we explain in appendix B this can be understood from the analyzing the in-falling boundary conditions at the special values of the frequency $\wn = -in$ with $n\in \mathbb{N}$. That both diffusion modes show such similarities leads one naturally to suspect that this might be a universal property of holographic hydrodynamic diffusion modes. At the moment it is not clear how such a conjectured universal behaviour could be proved. However it seems doable and an interesting task to generalize the calculations of the residues presented here and in \cite{Amado:2007yr} to other holographic gauge theories and check if the diffusive modes behave in the same way \cite{progress}. A clue of why this might happen comes from causality: in order to preserve causality and at the same time reproduce (first order) hydrodynamics in the long wavelength limit some drastic modification at short wavelengths is definitely necessary.

As we have seen, the complex momentum modes needed for the calculation of the front velocity do behave perfectly causal and their residues do not decouple at high frequencies. Although in the small frequency/momentum limit the lowest complex momentum mode can be computed from the analytic continuation of the quasinormal hydrodynamic mode $\wn = -i\qn^2$ this is not so for short wavelengths and high frequencies. This does not come as a surprise, since only rotational invariance is preserved in the finite temperature theory, so non-analyticities between $\omega$ and $q$ dependence are expected, like $\omega/q$. An example of this is the fact that the two limits $\omega\to 0$ and $q\to 0$ of the retarded correlator do not commute in general. Also, in order to do the analytic continuation of the Green function properly, all modes must be taken into account, only in the $\omega\to 0, q\to 0$ limit the hydrodynamic mode dominates. 
It is interesting to consider the hydrodynamic time scale we found in the energy density correlator, $2\pi T \tau_{\rm H} \approx 1.34 $. At RHIC temperatures $T\approx 300$ MeV this translates into a very short time $\tau_{\rm H} = 0.14$fm/c.\footnote{For the R-charge density this is even around four times shorter. This time has been erroneously reported to be $0.3$ fm/c in \cite{Amado:2007yr}.} One might take this as an indication for an extremely fast thermalization time. Of course, thermalization at RHIC includes processes that are outside the regime of linear response considered here, so the short hydrodynamic scale can at best describe a late stage of thermalization.

Another important point was to see how well the retarded Green's functions can be approximated by keeping only a few quasinormal modes. We found that analytic pieces related to the non-convergence of the sum over quasinormal modes played an important role. Recently, attempts of reconstructing the quasinormal mode spectrum from (much easier to compute) holographic spectral functions have been made in \cite{Myers:2008cj}. We think that our observations might also be useful to gain a better control on such procedures.

\begin{acknowledgments}
K.\,L. is supported by a Ram\'on y Cajal contract. S.\,M. is supported by an FPI 01/0728/2004 grant from Comunidad de Madrid. I.\,A. is supported by grant BES-2007-16830.
I.\,A., K.\,L. and S.\,M. are supported in part by the Plan Nacional de Altas Energ\'{\i}as FPA-2006-05485, FPA-2006-05423 and EC Commission under grant MRTN-CT-2004-005104. S.\,M. wants to thank G. S\'anchez for her support. We would like to thank G. Aarts, D. Kaplan, P. Kumar, E. L\'opez, I. Papadimitriou, A. Rebhan, A. Starinets and A. Vuorinen for useful discussions.
\end{acknowledgments}

\appendix

\section{\label{sec:method}Method}
\subsection{Energy-momentum tensor}
In the following we will use the coordinate $u=r_0^2/r^2$ such that the horizon sits at $u=1$ and the boundary at $u=0$. The equations of motion for the diffeomorphism-invariant quantities defined in \erf{eq:metriccomp} are \cite{Kovtun:2005ev}
\begin{subequations}
\begin{eqnarray}
 && Z_1''+\frac{(\wn^2-q^2 f(u)) f(u) -  u \wn^2 f'(u)}{u f(u) (\qn^2 f(u)-\wn^2)} \,Z_1' + \frac{\wn^2-\qn^2 f(u)}{u f(u)^2} \,Z_1  = 0 ~, \\
&& Z_2'' -\frac{3\wn^2(1+u^2)+\qn^2(2 u^2-3 u^4-3)}{ u f(u) (3 \wn^2+\qn^2(u^2-3)} \,Z_2' + \nonumber \\
&& \hspace*{3em} +\frac{3 \wn^4+\qn^4(3-4u^2+u^4)+\qn^2(4u^5-4u^3+4\wn^2 u^2 -6\wn^2)}{u f(u)^2(3\wn^2+\qn^2(u^2-3))} \,Z_2  = 0 ~, \qquad\qquad \\
&& Z_3''+\frac{1+u^2}{u f(u)} \,Z_3' +\frac{\wn-\qn^2 f(u)}{u f(u)^2} \,Z_3  = 0 ~.
\end{eqnarray}
\end{subequations}
Both the boundary and the horizon are regular singular points. At the horizon ($u=1$), there are two possible local solutions
\begin{equation}
Z_{(a)} \simeq (1-u)^{-i \wn/2} \,\varphi_{(a)}^{\rm in} + (1-u)^{i\wn /2} \,\varphi_{(a)}^{\rm out} ~, \quad a=1,2,3.
\end{equation}
In order to compute the retarded Green function, we must pick infalling boundary conditions $\varphi_{(a)}^{\rm out}\equiv 0$. According to the holographic dictionary the retarded Green function can be computed as the ratio of the connection coefficients that relate the local solution at the horizon with ingoing boundary conditions to the non-normalizable $(\cA_a)$ and normalizable $(\cB_a)$ solutions at the boundary ($u=0$)
\begin{equation}
Z_{(a)}^{\rm in} = \cA_{(a)} \varphi_{(a)}^1 + \cB_{(a)} u^2 \varphi_{(a)}^2 ~.
\end{equation}
The Green functions in the different channels are determined by three scalar functions $G_{(a)}$ given by the ratio of the connection coefficients
\begin{equation}
G_{(a)} = - \pi^2 N^2 T^4 \frac{\cB_{(a)}}{\cA_{(a)}}\,.
\end{equation}
The quasinormal modes are normalizable solutions where $\cA_{(a)}=0$. The ratio $\cB_{(a)}/\cA_{(a)}$ follows from demanding that the solution is smooth at a matching point in the interior of the interval $(0,1)$
\begin{equation}\label{eq:connection_coeffs1}
\frac{\cB_{(a)}}{\cA_{(a)}} = \frac{ Z_{(a)}^{\rm in} (Z_{(a)}^1)' - (Z_{(a)}^{\rm in})' Z_{(a)}^1}{ Z_{(a)}^2 (Z_{(a)}^{\rm in})' - (Z_{(a)}^2)' Z_{(a)}^{\rm in} } ~.
\end{equation}
We have computed the Frobenius series up to order 50. Matching the series expansions, we see that the ratio (\ref{eq:connection_coeffs1}) remains constant for a fair interval in the radial coordinate. We have chosen $x=0.53$ to evaluate the ratio.

The residue $R_n$ for the quasinormal mode $\wn_n$ can be computed as
\begin{equation}\label{eq:residue}
R_n = \left[ \left. \frac{\partial}{\partial \wn} \left(\frac{\cA_a}{\cB_a}\right)\right|_{\wn=\wn_n} \right]^{-1} ~.
\end{equation}

\subsection{Global current}
The equations of motion for the gauge invariant combinations (section \ref{sec:vector}) of the plane-wave vector field perturbations are \cite{Kovtun:2005ev}
\begin{subequations}\label{eq:vectorfields}
\begin{eqnarray}
E_T'' + \frac{f'(u)}{f(u)} \,E_T' + \frac{\wn^2- f(u) \qn^2}{(u f(u)^2} \,E_T &=&0 ~, \\
E_L'' + \frac{\wn^2 f'(u)}{f(u)(\wn^2 -f(u) \qn^2)} \,E_L' + \frac{\wn^2 - f(u) \qn^2}{u f(u)^2} \,E_L &=& 0 ~.
\end{eqnarray}
\end{subequations}
Defining $(\alpha):=(T,L)$ as the two gauge-invariant components components, we can follow the same procedure as with the energy-momentum tensor components. The infalling solution can be expanded at the boundary in the non-normalizable and normalizable modes
\begin{equation}
E_{(\alpha)}^{\rm in}(u) = \cA_{(\alpha)} \,E_{(\alpha)}^1(u) + \cB_{(\alpha)} \,E_{(\alpha)}^2(u) ~.
\end{equation}
where $E_{(\alpha)}^2(u)\sim u$. The retarded Green function is determined by the longitudinal and transverse polarization 
\begin{equation}
\Pi^{(\alpha)} = -\frac{N^2 T^2}{8} \frac{\cB_{(\alpha)}}{\cA_{(\alpha)}} 
\end{equation}
which are
proportional to the ratio between the connection coefficients
\begin{equation}\label{eq:connection_coeffs2}
\frac{\cB_{(\alpha)}}{\cA_{(\alpha)}} = \frac{ E_{(\alpha)}^{\rm in} (E_{(\alpha)}^1)' - (E_{(\alpha)}^{\rm in})' E_{(\alpha)}^1}{ E_{(\alpha)}^2 (E_{(\alpha)}^{\rm in})' - (E_{(\alpha)}^2)' E_{(\alpha)}^{\rm in} } ~.
\end{equation}
We have computed the Frobenius series up to order 50. Matching the series expansions, we see that the ratio (\ref{eq:connection_coeffs2}) remains constant for a fair interval in the radial coordinate. We have chosen $x=0.53$ to evaluate the ratio and have checked that the spectral function agrees with previous numerical (for non-zero momentum) and exact (for zero momentum) results \cite{Teaney:2006nc, Kovtun:2006pf, Myers:2007we}. The quasinormal modes correspond to the frequencies where there is a pole $\cA=0$. We can apply equation (\ref{eq:residue}) to compute the residues.

\section{\label{sec:zeroes}Zeroes of hydrodynamic residues}
We have seen that the residues of the diffusion and the shear mode have an oscillatory behaviour with the momentum. In this section we show how to find the location of the zeroes of the residues. We will use that the equations of motion for the for the vector field and vector component of the metric are Heun equations (c.f. \cite{Amado:2007pv, Hoyos:2006gb})\footnote{There is a factor of two difference with the conventions used here for the frequency and the momentum.}
\begin{equation}
y''(x) +\left( \frac{\gamma}{x} +\frac{\delta}{x-1}
 +\frac{\epsilon}{x-2} \right)y'(x) +\frac{\alpha\beta
  x-Q}{x(x-1)(x-2)}\,y(x) =0 ~.
\end{equation}

In the case of the vector field, we define $V_0(z)=A'_0(z)$, $V_L(z)=A'_L(z)$. In the $x=1-z^2$ coordinate, the Heun equations are found using the new variables
\begin{eqnarray}
V_0(x) & = &x^{-i \wn/2} (x-1)^{1/2} (x-2)^{-\wn/2} y(x) ~, \nonumber \\[-1ex] \\[-1ex] 
V_L(x) & = & x^{-1-i \wn/2} (x-1)^{1/2}(x-2)^{-1-\wn/2} y(x) ~. \nonumber
\end{eqnarray}
In both cases we find the same parameters for the Heun equation
\begin{eqnarray}\label{eq:longitudinalparameters}
\alpha&=& -\frac{\wn}{2}(1+i)~,\quad \beta= 2-\left(\frac{\wn}{2}(1+i)\right) ~,\quad Q =\qn^2 -(1+3i)\frac{\wn}{2} -(2-i)\frac{\wn^2}{2} ~, \qquad\nonumber \\[-1ex] \\[-1ex]
\gamma &=& 1-i\wn  ~,\quad \delta=1 ~,\quad \epsilon =1-\wn ~. \nonumber
\end{eqnarray} 
For the shear component we use the gauge-invariant variable $\psi_V$ proposed in \cite{Kodama:2003jz}. The Heun equation is found for the new variable
\begin{equation}
\psi_V(x)=x^{-i \wn/2} (x-1)^{3/4} (x-2)^{-\wn/2} y(x) ~,
\end{equation}
with parameters
\begin{eqnarray}
  \alpha\beta &=& \frac{\wn}{2}(1+i)\left( \frac{\wn}{2}(1+i)-3 \right)
  ~,\quad Q =\qn^2 -(1+5i)\frac{\wn}{2} -(2-i) \frac{\wn^2}{2} ~, \qquad
\nonumber \\[-1ex] \\[-1ex]
  \gamma &=& 1 -i\,\wn ~,\quad \delta=2 ~,\quad \epsilon=1-\wn ~. \nonumber
\end{eqnarray}
The coefficients of the Frobenius series at $x=0$ should satisfy the recursion relation
\begin{equation}\label{eq:heun.recursion}
 2(n+2)(n+1+\gamma) a_{n+2} +A_n(\wn,\qn) \,a_{n+1} +B_n(\wn,\qn) \,a_n=0 ~,\quad n\geq 0 ~,
\end{equation}
where 
\begin{eqnarray}\label{eq:ttrr}
A_n(\wn,\qn) &=& -((n+1)(2\delta+\epsilon+3(n+\gamma))+Q) ~, \quad \\
B_n(\wn,\qn) &=& (n+\alpha)(n+\beta) ~,
\end{eqnarray}
and $2\gamma a_1-Q a_0=0$. This recursion relation has in general a unique solution. However, when the continued fraction 
\begin{equation}\label{eq:continuedfraction}
  r_n =\frac{a_{n+1}}{a_n} = -\frac{B_n(\omega,q)}{A_n(\omega,q)+r_{n+1}} ~
\end{equation}
converges, Pincherle's theorem states that an extra solution to the three term recursion relation (\ref{eq:heun.recursion}) exists \cite{Leaver:1985ax, Birmingham:2001pj, Starinets:2002br}. This is equivalent to finding a solution that is analytic both at the boundary and at the horizon. Notice that at the horizon ($x=0$), the critical exponents are $0$ and $1-\gamma= i \wn$, so the analytic solution usually corresponds to the first solution, that is the one associated to infalling boundary conditions. However, when $\wn =-i k$, with $k=1,2,\dots$, the analytic solution will be in general the outgoing solution, since it corresponds to the largest integer critical exponent, while the infalling solution will have a logarithmic contribution. This is reflected in the recursion relations, since the coefficient of the $a_{n+2}$ term vanishes when $n+2=k$, implying that the series start at $x^k$. The reason for this is that the first $k$ recursion relations form a closed subsystem of linear equations for $k$ unknowns. Then, the general solution to the Heun equation will be of the form
\begin{equation}
y(x)= \alpha_1 x^k \sum_{n=0}^\infty  a_n x^n + \alpha_2 \left(\sum_{n=0}^\infty b_n x^n + c(\qn) \log(x) x^k \sum_{n=0}^\infty  a_n x^n \right) ~.
\end{equation} 
In principle, there could be special values of the momentum where the coefficient of the logarithm vanishes $c(\qn)=0$ and the two solutions at the horizon are analytic. This actually happens when the rank of the subsystem of the first $k$ recursion relations is zero. In that case, we can find trivially two solutions satisfying the recursion relations without having to check the convergence of (\ref{eq:continuedfraction}). 

The diffusion and shear modes are located at negative imaginary values of the frequency, larger as the momentum is increased. This implies that for some large enough value of the momentum, the frequency will have the special value $\wn=-i k$. However, the mode has to be analytic at the boundary and the horizon, so the value of the momentum when the special value is reached must be determined by the condition $c(\qn)=0$. When we compare the analytic result with the numerical computation, we find that this is indeed the case, the first points are $(i\wn ,\qn^2)=(1,1/2),\, (2,\sqrt{3}-1),\, (3,\sqrt{6}-3/2)$ for the diffusion mode and $(i\wn ,\qn^2)=(1,\sqrt{6}),\, (2,3.2266),\, (3,3.91764)$ for the shear mode. On the other hand, at the special point the analytic solution that is found as the limit of the hydro mode $\wn\to -i k$ is a linear combination of the normalizable and non-normalizable modes. This implies that the pole disappears from the retarded Green function or in other words, that the residue is zero.

\section{\label{sec:frontveloc}Front velocity}
Wave propagation in dispersive media has been studied long ago in the classic work of Brillouin and Sommerfeld \cite{Sommerfeld, Brillouin1, Brillouin2}. It has been pointed out there that the group velocity $v_g=\dd\omega /\dd q$ is not a reliable indicator if one wants to study the question of how fast can a signal be transmitted through the dispersive medium. In fact it is the so called front velocity which limits the speed of propagation of a signal through the medium. The front velocity is defined as the velocity with which the onset of a signal travels. In dispersive media this is not yet sufficient to guarantee that the signal travels with this speed and therefore one also has to define a signal velocity, which is the speed with which practically usable signals travel. This signal velocity is always smaller than the front velocity. For matters of principle, i.e. answering the question if causality is preserved it is therefore the front velocity that is important. We will briefly review here the reasoning leading to the definition of the front velocity. We want to study how fast a perturbation can travel through the medium, for this purpose we will switch on a periodic signal with frequency $\nu$ at time $t=0$. We model this by a source of the form $\Theta(t) \e^{ - i \nu t}\delta(x)$.\footnote{For simplicity we chose to localize the source in space, it is not important for the general argument.} To compute the response of the system we expand the retarded Green function in its poles in the complexified momentum plane (for simplicity we will also restrict our considerations to effectively $1+1$ dimensions)
\begin{equation}
\vev{\Phi(x,t)} = - \int \frac{\dd\omega}{2\pi}\frac{\dd q}{2\pi} \sum_n \frac{ R_n(\omega,q)}{q-q_n} \frac{i}{\omega-\nu + i\epsilon} \,\e^{-i \omega t + i q x} ~.
\end{equation}
The poles in the momentum $k$ have real and imaginary parts $q_n(\omega) = q^R_n(\omega)+ i q^I_n(\omega)$ and from \cite{Amado:2007pv, Hoyos:2006gb} we know that they have to lie symmetrically in the first and third quadrants. We consider therefore only the region with $x>0$ since the response in $x<0$ is just the mirror image. Assuming $x>0$ we have
\begin{equation}
\vev{\Phi(x,t)} = - \int \frac{\dd\omega}{2\pi} \sum_n R_n(\omega,q_n) \,\frac{i}{\omega-\nu + i\epsilon} \,\e^{-i \omega t + i q^R_n x} \,\e^{-q^I_n x} ~.
\end{equation}
To pick up a non vanishing signal we have to close the contour of integration now in the lower half plane. This demands however that $\lim_{\omega\rightarrow \infty}( t - x \frac{q^R_n(\omega)}{\omega} ) \geq 0$ showing that we can pick up the first front of the signal only at space time points where $x/t \leq v_{\rm F}$, where we have defined the front velocity\,\footnote{Since this implies a linear relation between $\omega$ and $q$ for large frequencies and (real) wave numbers it is usually written as $v_{\rm F} =\lim_{q\rightarrow\infty} \frac{\omega}{q}$. It also follows that the diffusion equation violates causality since from the dispersion relation we find $\omega=2 D q^2$ and therefore $\omega/q = 2D q$ with no finite limit for the front velocity.}
\begin{equation}
v_{\rm F} := \lim_{\omega\rightarrow\infty} \frac{\omega}{q^R} ~.
\end{equation}
It follows that causality is preserved if the front velocity is smaller that the speed of light $v_{\rm F} \leq c$. Results for the complex momentum modes for the shear channel and the diffusion channel in \cite{Amado:2007pv, Hoyos:2006gb} and for the sound channel in figure \ref{fig:CMMsound}, show that the imaginary parts go to zero if we extrapolate to the large frequency limit, i.e. $\omega\to\infty$, while the real part approaches from below the dispersion relation $\omega=q$. Moreover, the evolution of the hydrodynamic modes in this limit is the same as the evolution of the higher modes, finding numerically that $\lim_{\omega\rightarrow\infty} \frac{\omega}{q}=1$. This provides a numerical proof that the strongly coupled $\Nfour$ plasma behaves causally even when we take into account the hydrodynamic modes.

\bibliographystyle{unsrt}
\bibliography{HydroScales}
\end{document}